\documentclass[manuscript,11pt]{aastex}
\usepackage{epsfig}
\usepackage{soul}
\usepackage{color}

\usepackage{physymb}
\usepackage{mathrsfs}

\usepackage{lineno}

\def\sxvi{S\,{\sc xvi}}

\def\sixiv{Si\,{\sc xiv}}

\def\mgxii{Mg\,{\sc xii}}
\def\nex{Ne\,{\sc x}}

\def\civ{C\,{\sc iv}}

\def\mgxii{Mg\,{\sc xii}}

\def\civ{C\,{\sc iv}}

\def\nv{N\,{\sc v}}

\def\mathv{\textbf{\em v}}

\def\mathn{\textbf{\em n}}

\def\cm{\ifmmode {\rm cm}^{-1} \else cm$^{-1}$ \fi}
\def\s{\ifmmode {\rm s}^{-1} \else s$^{-1}$ \fi}
\def\cc{\ifmmode {\rm cm}^{-3} \else cm$^{-3}$ \fi}
\def\cs{\ifmmode {\rm cm}^{-2} \else cm$^{-2}$ \fi}
\def\g{\ifmmode \gamma \else $\gamma$\fi}
\def\G{\ifmmode \Gamma \else $\Gamma$\fi}
\def\Gs{\ifmmode \Gamma~ \else $\Gamma~$\fi}

\def\gc{\ifmmode \gamma_{\rm c} \else $\gamma_{\rm c}$ \fi}
\def\sw{Schwarzschild~}
\def\gsim{\mathrel{\raise.5ex\hbox{$>$}\mkern-14mu
             \lower0.6ex\hbox{$\sim$}}}
\def\lsim{\mathrel{\raise.3ex\hbox{$<$}\mkern-14mu
             \lower0.6ex\hbox{$\sim$}}}
\def\simless{\mathbin{\lower 3pt\hbox
     {$\rlap{\raise 5pt\hbox{$\char'074$}}\mathchar"7218$}}}   
\def\simmore{\mathbin{\lower 3pt\hbox
     {$\rlap{\raise 5pt\hbox{$\char'076$}}\mathchar"7218$}}}   
\def\Msun{M_\odot}                                
\def\deg{^\circ}

\def\gro1655{GRO~J1655-40}
\def\4u1630{4U1630-472}
\def\h1743{H1743-322}
\def\grs1915{GRS1915+105}

\shorttitle{Wind Scattering}

\shortauthors{Fukumura 2025}

\begin{document}


\title{Compton Scattering of Thermal Disk Radiation with Black Hole Disk Winds}


\date{\today}

\author{
\textsc{Keigo Fukumura}\altaffilmark{1}
}

\altaffiltext{1}{Department of Physics and Astronomy, James Madison University,
Harrisonburg, VA 22807; fukumukx@jmu.edu}

%
%
%
%
%
%
%
%
%

\begin{abstract}
\baselineskip=15pt

Galactic black hole (BH) X-ray binaries are known to exhibit episodic outbursts, during which accretion and spectral mode distinctively transition between low/hard and high/soft state. X-ray observations during high/soft state occasionally reveal a pronounced presence of a powerful disk wind in these systems. However, it is  unexplored to date how  such winds  may influence  disk emission in that regime. 
In this work, we consider an observational implication by Compton scattering of thermal disk radiation due to  accretion disk winds by performing multi-dimensional Monte Carlo simulations in the context of a stratified  wind of large solid angle  launched over a large radial extent of the disk.  Compton-scattered thermal disk spectrum is computed for a different wind property; i.e. wind density and its radial gradient. We find that the intrinsic  disk radiation can be significantly down-scattered by winds of moderate-to-high density to the extent that the transmitted spectrum can substantially deviate from the  conventional multi-color-disk emission  in a tangible way. 
We thus claim that the conventional treatment of spectral hardening in the disk atmosphere may  be insufficient to fully account for the observed disk continuum in the presence of strong wind scattering.  
It is  suggested that the effect of scattering process (by $f_w$) should be  incorporated to accurately evaluate an intrinsic disk spectrum besides the conventional hardening (color correction) factor (by $f_c$). We argue that BH spin measurements using thermal continuum-fitting in transient XRBs may well be mildly (if not significantly) altered by such spectral ``contamination".


\end{abstract}


\keywords{Accretion (14) --- Computational methods (1965) --- Low-mass x-ray binary stars (939) --- 
Stellar mass black holes (1611) --- X-ray transient sources (1852)  --- Black hole physics (159) --- Spectroscopy (1558) 
--- Astrophysical black holes(98) --- X-ray astronomy(1810)}



\baselineskip=15pt

\section{Introduction}

Transient galactic sources, such as accretion-powered black hole (BH) X-ray binaries (XRBs), are well known to exhibit a unique hysteresis pattern in intensity and hardness relation during an episodic outburst phase as the system transitions into different mass accretion modes with varying X-ray brightness \citep[e.g.][]{Esin97,Done07,Ponti12}.  It is generally seen during each outburst that the observed X-ray spectrum  changes between Comptonized state of power-law emission and thermally-dominant state of blackbody radiation   as accretion state changes between low-intensity/hard-spectrum (i.e. low/hard) state and high-intensity/soft-spectrum (i.e. high/soft) state \citep[e.g.][]{Esin97,Done07,Ponti12}. It is of particular interest from an extensive number of X-ray spectroscopy that high/soft state of BH XRBs is occasionally accompanied by powerful outflows from accretion disks in the form of blueshifted absorption features from a series of highly ionized ions\footnote[1]{This is not exclusive to BH XRBs, but also seen in neutron star low-mass XRBs; e.g. GX~13+1 \citep[e.g.][]{Ueda04}.} \citep[e.g.][]{Kallman09,Miller12,Miller15}, while mostly undetected during low/hard state\footnote[2]{It is found, however, that optical winds are in fact present almost exclusively during low/hard state \citep[e.g.][]{MunozDarias19,Mata22}, while InfraRed winds appear to be persistent throughout the outburst in these transients \citep[e.g.][]{SanchezSierras20}. In this work, we are focused on X-ray winds for a coherent study.}. Hence,  prominent disk winds normally coexist with dominant disk radiation when the system shows high mass-accretion rate.  

Within the framework of the standard accretion disk model of  a geometrically-thin, optically-thick configuration \citep[][]{SS73,NT73} during high/soft state in BH XRBs, it has been conventionally conceived that the  thin disk is smoothly covered by a layer of hot atmosphere in a vertical structure \citep[e.g.][]{ShimuraTakahara93,Shimura95} in which soft thermal X-ray photons from the disk are subject to  scattering and absorption. In a classical work by \citet[][hereafter ST95]{ShimuraTakahara95},  vertical radiative transfer has been numerically solved  above the  disk surface to determine emergent thermal spectrum. ST95 discuss that the emergent continuum spectrum from the disk can be approximately described as a {\it diluted} blackbody form modified by a {\it spectral hardening} factor (aka. {\it color correction} factor) of $f_c \sim 1.7-2$ depending on the Eddington ratio and disk viscosity parameter. Later, their pioneering work was re-examined  
by utilizing the {\tt TLUSTY} stellar atmospheres code \citep[][]{Hubeny90, HubenyLanz95} to compute models of the vertical structure and radiation transfer in the accretion disk annuli \citep[e.g.][]{Davis05,Davis19} in which $f_c \lsim  1.7-1.9$ is usually obtained.

\begin{figure}[t]
\begin{center}
\includegraphics[trim=0in -0.5in 0in
0in,keepaspectratio=false,width=3.6in,angle=-0,clip=false]{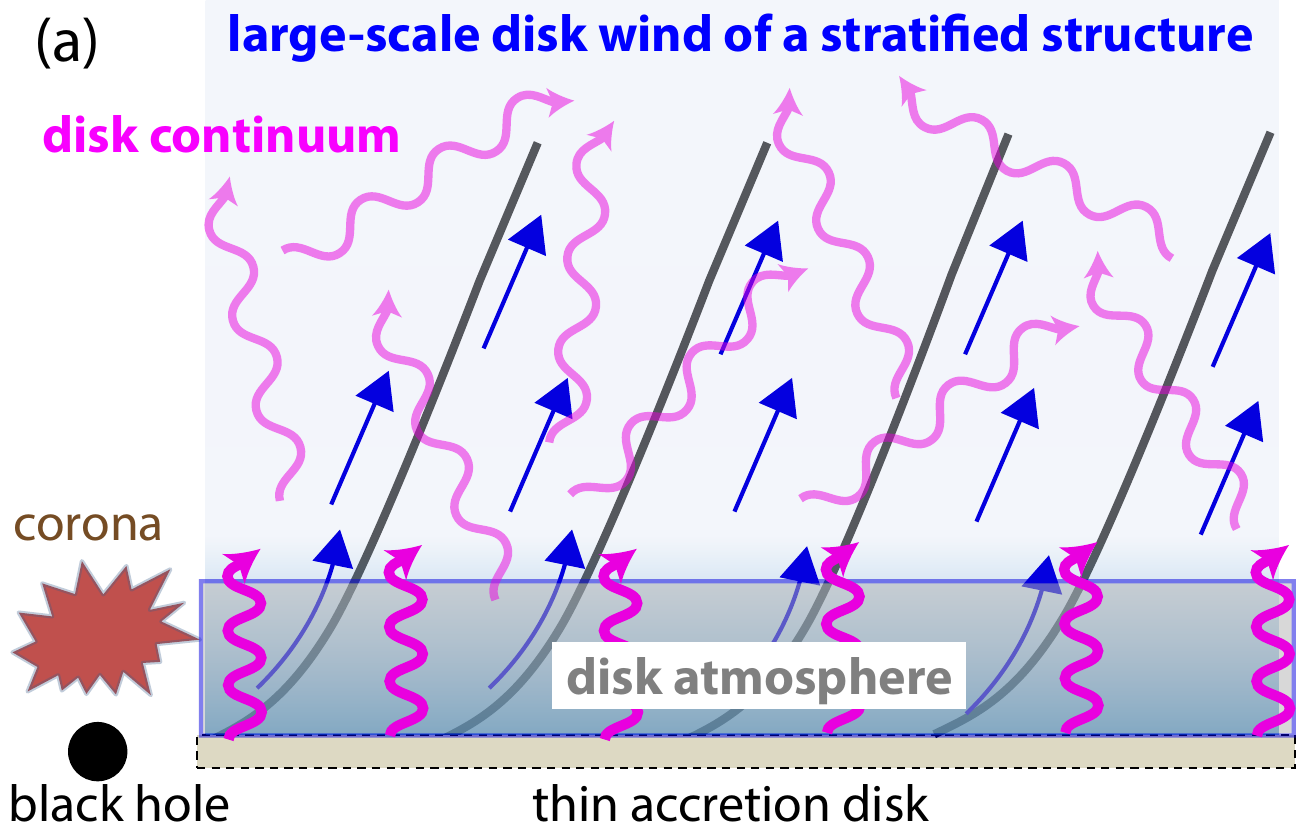}\includegraphics[trim=0in -0.5in 0in
0in,keepaspectratio=false,width=2.4in,angle=-0,clip=false]{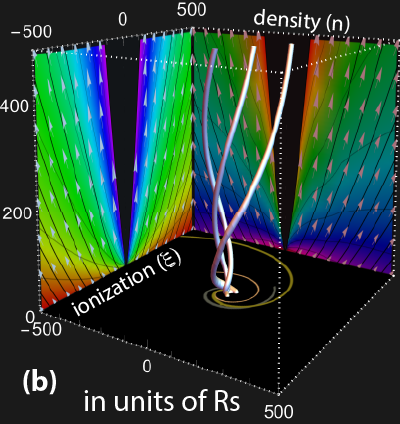}
\end{center}
\caption{(a) Schematic diagram to illustrate scattering of accretion disk thermal continuum  with  large-scale disk wind of  stratified geometry. In this work, we ignore Comptonized hardening by the disk atmosphere \citep[e.g.][]{ShimuraTakahara95} for simplicity. (b) Close-up 3D rendered streamlines of a simulated wind for $p = 1.3$ launched from a thin disk surface. Poloidal distribution of normalized wind density $\log (n/n_{in})$ and ionization parameter $\log(\xi/\xi_{in})$ are shown in color superimposed by wind velocity vector $v(r,\theta)$ (white arrows) and poloidal projection of the magnetic field lines (thick dark solid) in the innermost wind domain for $-500 \le x/R_S, y/R_S \le 500$ (from \citealt{F18}). }
\label{fig:cartoon}
\end{figure}

Meanwhile, thermally-dominant disk spectrum in BH XRBs have also been proven to be  an effective utility to constrain BH spin since the continuum spectrum during high/soft state (where winds are also present) is uniquely sensitive to the radius of the innermost stable circular orbit (ISCO), which is an explicit function of BH spin \citep[e.g.][]{Bardeen72}. By carefully analyzing the observed thermal-disk spectra, a number of spin  measurements have been made based on this {\it continuum-fitting} method \citep[e.g.][]{Davis05, Li05,Shafee06,McClintock06,Steiner09,McClintock14} as one of a few techniques available. In this method, {\it spectral hardening} of the disk radiation through the disk atmosphere is built in for spectral analysis by adopting a fiducial value of the {\it color correction} factor of $f_c = 1.7$ following ST95. However, despite the persistent presence of strong disk winds when the continuum-fitting method is applied, the physical impact of disk winds on emergent thermal spectrum has never been explicitly addressed or  quantitatively investigated in a tangible fashion to date. Motivated by the lack of such a perspective, we argue, for the first time, that such a disk wind  could {\it down-scatter} diluted disk photons modifying the expected emergent spectrum. The essence of our model is schematically illustrated in {\bf Figure~\ref{fig:cartoon}a} where thermal blackbody radiation locally dissipated over a large radial extent is Compton down-scattered by a large-scale disk winds of a stratified structure. In the context of BH XRBs undergoing disk-dominated spectral state (i.e. high/soft accretion mode), the intrinsic thermal continuum, often characterized by the multi-color-disk (MCD) emission \citep[][]{Mitsuda84}, is expected to be modified as a result. To simplify the problem, we ignore in this work the effect of spectral hardening through the disk atmosphere (ST95).    
Depending on the wind condition including geometry, wind density and its radial profile, we find in this work that a tangible effect of scattering  could potentially alter the  emergent spectrum, potentially leading to different estimate of BH spin using the continuum-fitting method.

In \S 2, we briefly explain the primary foundation of the  wind models of a characteristic structure. In \S 3, we present our results demonstrating the expected disk spectrum taking into account scattering with disk winds. We summarize our findings and discuss the implications of our model  in \S 4, followed by Appendices A-C where a set of supplementary materials are presented for comparison.

\section{Model Description}

In this work, we consider  a continuous wind launched over a large radial extent of an accretion disk  by utilizing the existing theoretical framework of magnetohydrodynamic (MHD) disk wind model (i.e.. \citealt{BP82}; \citealt{CL94})
%
%
of certain stratification in wind properties\footnote[3]{Note that the most essential feature of our wind model in this paper is its stratification in density and velocity, for example. Hence, the exact type of driving mechanism, while intriguing,  is not critically important here. }, which has already been shown to successfully explain diverse wind absorption phenomena in both AGNs and BH XRBs; e.g. X-ray outflows (\citealt{F10a}, hereafter F10; \citealt{Chakravorty16,Gandhi22,Chakravorty23}; \citealt{F17}; \citealt{F18}; \citealt{F21}), ultra-fast outflows \citep[][]{F15,Kraemer18,F22}, and broad UV \civ\ features \citep[][]{F10b} as well as the observed obscuration in X-ray and broad \civ\ absorption in UV often seen in Seyfert 1s  \citep{F24}.

\subsection{Disk Wind Model of Stratified Structure}

A key nature of the stratified wind considered here is mainly characterized  by its smooth distribution in density and velocity
\begin{eqnarray}
n(r,\theta) =  n_o \left(\frac{r}{R_o} \right)^{-p} f(\theta)  \ ,   \label{eq:density}
\end{eqnarray}
such that the wind density radially falls off in a self-similar fashion determined by the index $p>0$ while also varying in angular direction governed by a function $f(\theta)$ which is numerically obtained by solving the ideal MHD equations coupled to Grad-Shafranov equation (see F10) under steady-state, axisymmetric assumptions. In this formalism, therefore, the wind has the highest density on the disk surface (where $f(90\deg)=1$ by definition) at the innermost launching radius $R_o$, which is assumed to be effectively the ISCO radius. Thus, the 3D density  distribution (with axisymmetry) is completely dictated by specifying $p$ and a normalization value of $n_o$ where $n_o$ corresponds to the wind density at its base\footnote[4]{Therefore, $n_o$, which is observationally inaccessible in a direct way, is not to be confused with the absorber's density, $n$, derived from X-ray observations. See {\bf Figure~\ref{fig:cartoon}b}.} with $(r,\theta)=(R_o,90\deg)$. 
{\bf Figure~\ref{fig:cartoon}b} shows close-up 3D rendered streamlines of a simulated wind for $p = 1.3$ \citep[][]{F18} with the color-coded   density distribution in the poloidal plane $n(r,\theta)$ (higher in purple and lower in red). In addition, ionization parameter $\xi \propto 1/[n(r,\theta) r^2]$ of the wind is   also shown as a reference in the other poloidal plane (higher in purple and lower in red).

The kinematics of stratified wind is similarly described by
\begin{eqnarray}
\mathv(r,\theta,\phi) = (v_r, v_\theta, v_\phi)  \ ,   \label{eq:velocity}
\end{eqnarray}
and we have, for example, $v_r(r,\theta) = v_K(r) g(\theta)$  where $v_K(r)$ is the Keplerian (escape) velocity with the angular dependence given by a numerical solution $g(\theta)$ where $g(90\deg)=1$ by definition. Again, this is demonstrated in {\bf Figure~\ref{fig:cartoon}b} where the poloidal velocity (white arrows) is increasing along field lines while decreasing like $|\mathv| \propto r^{-1/2}$ along a line of sight (LoS).  
A large-scale morphology of the wind is determined by a number of conserved quantities of outflowing plasma; e.g. particle-flux to magnetic-flux, total angular momentum of plasma particle, and Bernouilli parameter, among others (\citealt{CL94} and F10 for details).    
To mitigate a complexity due to an additional degree of freedom in the model, we exploit in this work a fiducial wind geometry presented in {\bf Figure~\ref{fig:cartoon}b}  \citep[adopted from][]{F18}  such that we are focused specifically on examining the role of the wind density distribution given by  equation~(\ref{eq:density})  through scattering process. 
%
%

\begin{figure}[t]
\begin{center}
\includegraphics[trim=0in 0in 0in
0in,keepaspectratio=false,width=3.3in,angle=-0,clip=false]{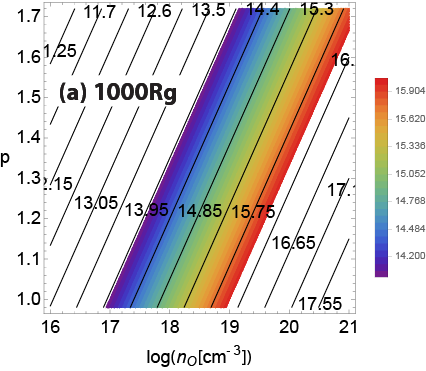}\includegraphics[trim=0in 0in 0in
0in,keepaspectratio=false,width=3.3in,angle=-0,clip=false]{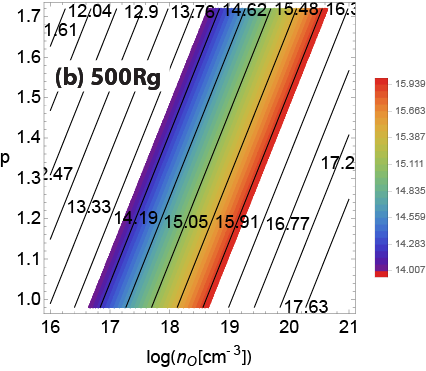}
\end{center}
\caption{Domain of the observed wind density, $14 ~ \rm{(blue)} \lsim \log (n \rm{[cm^{-3}]}) \lsim 16 ~\rm{(red)}$, in  ($n_o, p$)-parameter space inferred observationally from X-ray disk wind analyses \citep[e.g.][]{Kallman09,Miller08,Miller15,Trueba19,Ratheesh21,F21} at the distance  of (a) $r \simeq 1,000R_g$  and (b) $500R_g$. }
\label{fig:cont}
\end{figure}

To better probe an observationally plausible range of $(p, n_o)$ for our calculations, we make a reference to a number of robust spectral analyses in literature for BH XRB winds such as 4U~1630-47, GRO~J1655-40, H~1743-322, and GRS~1915+105, as an exemplary sample \citep[e.g.][]{Miller08,Kallman09,Miller15,Trueba19,Ratheesh21,F21} where multi-ion absorbers at different charge state have been detected. While some absorbers are most likely classified as Ly$\alpha$ or He$\alpha$ Fe K ions in these sources, some are robustly identified as low ionization absorbers in soft energy band such as \sxvi, \sixiv, \mgxii, \nex, and \nv, among others.  Based on their analyses using photoionization models in those studies, it is suggested  that X-ray winds can be conservatively characterized  by density range of $n \sim 10^{14-16}$ cm$^{-3}$ at distance range of $r/R_g \sim 500-1,000$. With equation~(\ref{eq:density}), one can then determine the range corresponding to $14 \lsim \log(n\rm{[cm^{-3}]}) \lsim 16$  in the parameter space of $(n_o,p)$ when the absorbers are located at (a) $r \simeq 1,000R_g$  and (b) $\sim 500R_g$, as shown in {\bf Figure~\ref{fig:cont}}. As seen, $n_o$ needs to be higher with larger $p$ to compensate for a rapid decline of wind density. Furthermore, $n_o$ is expected to be higher for a given $p$ when the observed wind location is further out. 
In parallel, the observed multiple absorbers of different ions can be globally accounted for in the form of stratified winds\footnote[5]{This could also be true and relevant for AGN winds as observed by independent studies \citep[e.g.][]{B09,Detmers11,Kosec18,F18,Yamada24}.} if the radial wind density structure has $p \sim 1-1.4$  \citep[e.g.][]{Trueba19,Ratheesh21,F21}.  
Given the obtained constraint from observations, therefore, we explore with the present model a restricted range of density normalization $10^{17} \lsim n_o \rm{[cm^{-3}]} \lsim 10^{21}$ for different density slope $1 \lsim p \lsim 1.7$ to cover {\tt case 1A-1C}, {\tt case 2A-2C} and {\tt case 3A-3C} as listed in {\bf Table~\ref{tab:tab1}}. We shall discuss this later in \S 3. 

\begin{deluxetable}{cl||l|c}
\tabletypesize{\small} \tablecaption{Stratified Disk Wind Parameters in the Model} \tablewidth{0pt}
\tablehead{Case  & & Parameters  &  \hspace{-0cm} {\bf A}  \hspace{1cm} {\bf B} \hspace{1cm} {\bf C} 
} 
\startdata
%
\hline \hline
{\bf 1} & Slow density & $p=1$   &     \\
   & decay in radius & $n_{o}$ [cm$^{-3}$]   &  \hspace*{0.3cm}  $10^{17}  \hspace*{0.8cm} 10^{18}  \hspace*{0.5cm}   5 \times 10^{18}$   \\  \hline
{\bf 2} & Moderate density & $p=1.2$   &     \\
   & decay in radius & $n_{o}$ [cm$^{-3}$]   &  \hspace*{0.3cm}  $10^{18}  \hspace*{0.8cm}  10^{19}   \hspace*{0.5cm}  5 \times 10^{19}$    \\  \hline
{\bf 3} & Fast density & $p=1.7$   & \\
   & decay in radius & $n_{o}$ [cm$^{-3}$]   &  \hspace*{0.3cm}  $10^{19}   \hspace*{0.8cm}  10^{20}  \hspace*{0.5cm}  5 \times 10^{20}$    \\  \hline
   \label{tab:tab1}
\enddata
\vspace*{0.2cm}
\end{deluxetable}


\subsection{Compton Scattering Process}

For a given wind property of a certain morphology as described in \S 2.1, we consider the effect of scattering on disk photons by carrying out 3D Monte Carlo (MC) simulations assuming a photon source on the disk at radius $r$. The mean-free-path of a photon is computed by     
\begin{eqnarray}
\lambda_{\rm mfp} \equiv \frac{1}{n(r,\theta) \sigma_{\rm KN}} \ ,
\end{eqnarray}
which depends on the scattering cross-section $\sigma_{\rm KN}$ from Klein-Nishina formula and local wind density $n(r,\theta)$. 


Following the standard MC method for scattering, we consider a scattering probability by $\xi \equiv 1-e^{-\tau}$ where the wind optical depth is given by $\tau \equiv L/\lambda_{\rm mfp}$ with photon's traveling distance $L$. Randomly sampling the value of $0 \le \xi \le 1$ for each scattering event at a given wind location,  $L$ is obtained in each step.




Because a pre-scattering emission direction (of an incoming photon) is different for each scattering event, we calculate a scattering direction of the photon with respect to the previous photon direction by using scattering angle $\psi$ and the azimuthal angle $\chi$ measured in the lab frame. From a geometrical relation, one finds that a pre-scattering photon propagating with a unit vector of $\tilde{\mathn}_o = (n_x, n_y, n_z)=(\tilde{\ell}_o, \tilde{m}_o, \tilde{n}_o)$ is going to be scattered into a new direction with a unit vector of  $\tilde{\mathn} = (n_x, n_y, n_z)=(\tilde{\ell}, \tilde{m}, \tilde{n}$) such that
\begin{eqnarray}
\tilde{\ell} &=& \tilde{\ell}_{o} \cos \psi_{\rm} +(\tilde{\ell}_{o} \tilde{n}_{o} \sin \psi \cos \chi - \tilde{m}_o \sin \psi \sin \chi)/\tilde{s}_o \ , \\
\tilde{m}_{\rm} &=& \tilde{m}_{o} \cos \psi +(\tilde{m}_o \tilde{n}_o \sin \psi \cos \chi + \tilde{\ell}_o \sin \psi \sin \chi)/\tilde{s}_o \ , \\
\tilde{n}_{\rm} &=& \tilde{n}_{o} \cos \psi - \tilde{s}_o \sin \psi \cos \chi \ , 
\end{eqnarray}
where $\tilde{s}_o = \sqrt{1-\tilde{n}_o^2}$ \citep[e.g.][]{Uehara84}. 
Again, the subscript ``o" denotes its position prior to scattering, $\psi$ is the scattering angle, and  $\chi$ is the randomly-sampled azimuthal scattering angle measured from its pre-scattering null direction. 
Hence, ($\tilde{\ell}, \tilde{m}, \tilde{n}$) will be iteratively replaced by  ($\tilde{\ell}_o, \tilde{m}_o, \tilde{n}_o$) for each scattering event. This is how we keep track of the history of null trajectories in MC calculations.

our MHD wind can be (near-)relativistic in terms of velocity (e.g. $v/c \lsim 0.7$) especially at smaller launching radius near BH due to Keplerian motion. Thus, the assumption of low-velocity or stationary electrons with respect to disk photons does not always hold. To correctly account for scattering direction and post-scattering energy shift of such high-velocity electrons, I decided to implement SR aberration for completeness. Since the wind (electron) velocity is indeed sub-relativistic elsewhere, one could also argue that the effect of SR aberration is only limited to a small spatial domain, most likely yielding a minimum influence to  the end results (i.e. scattered MCD spectrum). 

The wind medium (i.e. electrons) in our scheme is generally sub-relativistic in almost every part of the computational domain except for the innermost region and near the disk surface; i.e. $r/R_S \lsim 10$ and $\theta \sim 90\deg$. 
Hence, the assumption of low-velocity or stationary electrons (i.e. $v/c \ll 1$) with respect to disk photons does not always hold. 
We thus implement special relativistic (SR) aberration by the standard  transformation between the wind/electron rest frame and lab frame to properly handle very fast wind layer in relativistic regime (e.g. $v/c \lsim 0.7$) near the innermost part of the wind by computing aberration factor as
\begin{eqnarray}
\psi(\beta) = \rm{mod} \left[ \cos^{-1} \left \{\ \frac{\cos(\psi') + \beta}{\beta \cos(\psi') + 1}  \right \}\ , \pi \right] \ ,
\end{eqnarray}
where $\psi'$ measures photon's emission angle between the previous propagation direction and the scattering direction in the lab frame, $\beta \equiv v_{\rm proj}/c$  denotes the wind velocity component projected onto the photon propagation direction. $\psi$ is then the corresponding scattering angle in the wind/electron  rest-frame. The wind velocity $\mathv(r,\theta,\phi)$ from equation~(\ref{eq:velocity})
is  used to  determine $v_{\rm proj}$. 
Since SR aberration effect is limited to only small radius, we note that  the effects don’t matter practically to the primary results here. To further clarify, neither relativistic beaming from the fast rotating disk surface nor  gravitational redshift (energy shift) is  included in our calculations because these effects are restricted only to small radius near BH.

\begin{figure}[t]
\begin{center}
\includegraphics[trim=0in 0in 0in
0in,keepaspectratio=false,width=3.6in,angle=-0,clip=false]{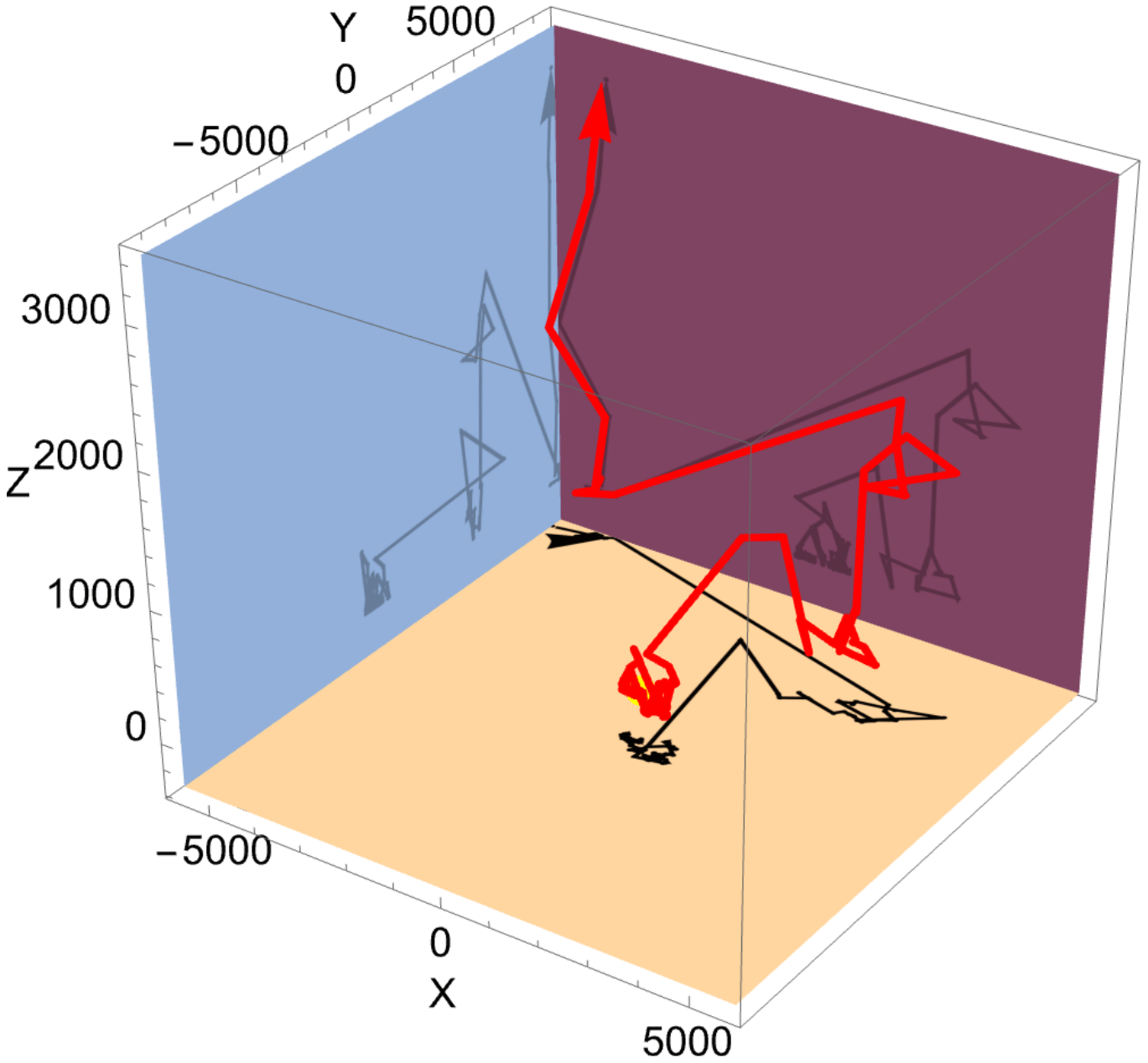}
\end{center}
\caption{Trajectory of a scattered photon initially emitted at $R=(X^2+Y^2)^{1/2}=Z=10R_g$ and $Z=5R_g$ in MC simulations through winds with $p=1.7$ and $n_{o}=10^{19}$ ({\tt case 3A}). Black hole is situated at the origin surrounded by a thin disk in the equator (i.e. $Z/R_g \sim 0$). {\bf This figure is available as an animation.} }
\label{fig:path}
\end{figure}

The post-scattering (i.e. final)  photon energy $E_f$ after each scattering event is given as usual by
\begin{eqnarray}
E_f = \frac{E_i}{1+y_i (1-\cos \psi')} \ , 
\end{eqnarray}
where $E_i$ is the pre-scattering (i.e. initial) photon energy and $y_i \equiv E_i/(m_e c^2)$ with $m_e$ and $c$ being the electron mass and the speed of light, respectively.

With this computational formalism, a subset of MC simulation is made visualized, as an example, to keep track of a scattered photon in {\bf Figure~\ref{fig:path}} for winds with $p=1.7$ and $n_o=10^{19}$ cm$^{-3}$ ({\tt case 3A}) where  a photon is emitted from $R=(X^2+Y^2)^{1/2}=10R_g$ and $Z=5R_g$.
{\bf This figure is available as an animation.}
It is found, as a generic behavior in all MC simulations, that many photons spend more time within the inner layer of the wind being repeatedly scattered around before they manage to escape to infinity due to  higher density. This trend becomes more pronounced with smaller $p$ and higher $n_o$ values as expected. In the presence of sub-relativistic disk wind in the current framework, photons loose energy repeatedly through Compton down-scattering. As discussed next in \S 2.3, this process naturally induces a consequential impact on a global characteristics of disk emission.

\subsection{Scattered Radiation from Thermal Accretion Disk}

We consider thermal blackbody continuum from a geometrically-thin, optically-thick accretion disk around a galactic black hole of mass $M=10\Msun$ in XRB during high/soft state  \citep[e.g.][]{Done07} where  strong disk wind is occasionally observed \citep[e.g.][]{Miller15}. 
The blackbody flux of local disk radiation produced at radius $r$ is given by
\begin{eqnarray}
B_E(r) \propto E^2 \left\{ e^{E/\left(k_B T_o \right)  \left(r/R_{\rm in}\right)^{3/4} } -1 \right\}^{-1} ~~ \rm{[photons~s^{-1}~cm^{-2}~keV^{-1}~str^{-1}]} \ ,
\end{eqnarray}
where the local  disk temperature follows $T(r) \propto r^{-3/4}$ in the standard thin disk theory \citep[e.g.][]{SS73,NT73} with the inner disk radius   $R_{\rm in} \equiv 3 R_S = 6 R_g$ being the ISCO radius of the disk around a \sw BH. The characteristic  temperature $T_o$ is given  at $r=R_{\rm in}$  and $k_B$  is Boltzmann constant. Based on X-ray observations of a number of BH XRBs during high/soft state, we assume here $k_B T_o=1$ keV as a canonical value \citep[e.g.][]{Kallman09,Miller15}. The exact choice of $T_o$ in this work does not qualitatively change the end results.  

We then make use of multi-dimensional MC ray-tracing to further compute Comptonized thermal spectrum at a specific radius. Hence, the scattered spectrum is given by  
\begin{eqnarray}
B^{\rm scat}_E(r) \equiv  B_E(r) \phi_{\rm scat}(r, E; p, n_o) \ ,
\end{eqnarray}
where $\phi_{\rm scat}(r,E; p,n_o)$ is a scattering kernel as a function of $(p,n_o)$ determining the degree of scattering obtained from the MC simulations when a point-like X-ray source is placed at radius $r$ (e.g., see {\bf Fig.~\ref{fig:path}}). Since  $B^{\rm scat}_E(r)$ is the differential flux per radius, the total spectrum to be observed is found by integrating over the whole radial extent of the disk from $r=R_{\rm in}$ to $r=R_{\rm out}$ as 
\begin{eqnarray}
S^{\rm scat}_E \propto \int_{R_{\rm in}}^{R_{\rm out}}  B^{\rm scat}_E(r) r dr   ~~ \rm{[photons~s^{-1}~cm^{-2}~keV^{-1}]}  \ ,
\end{eqnarray}
where we consider in this work $R_{\rm in} = 3R_S$ and $R_{\rm out} \simeq 10^3 R_S$ with a fixed aspect ratio of $Z/R=0.1$ to mimic a thin disk geometry\footnote[6]{We have tried a number of slightly different aspect ratios other than $0.1$ and find that the end results are qualitatively insensitive to a particular choice.}. In radial direction, the disk region is divided logarithmically into roughly 26 annuli in a way similar to the approach adopted in  ST95. Disk photons emitted in each annular are subject to different magnitude of scattering due to the spatial dependence of the wind property, namely, density and velocity profiles, as expressed in equations~(\ref{eq:density})-(\ref{eq:velocity}).  

While we perform MC simulations to numerically calculate disk spectrum that is down-scattered by disk winds, one can search for a convenient correction factor, $f_{w,i}$, to account for this effect on the intrinsic local disk spectrum such that
\begin{eqnarray}
B^{\rm scat}_E(T)  \propto \frac{1}{f_{w,1}^4}   B_E(f_{w,2} T) + \frac{1}{f_{w,3}^4}   B_E(f_{w,4} T)   \ , 
\label{eq:bestfit}
\end{eqnarray}
as a superposition of two independent blackbody components, which turns out to be phenomenologically necessary to properly fit some of the extremely Comptonized disk spectra as shall be discussed in \S 3 (thus a single parameterization with $f_w$ in a single blackbody form would not work). The role of {\it scattering} is therefore manifested in these correction factors $f_{w,i}$ ($i=1,2,3,4$) just like the role of {\it hot disk atmosphere} is quantified by hardening factor $f_c$ in ST95.

\begin{figure}[t]
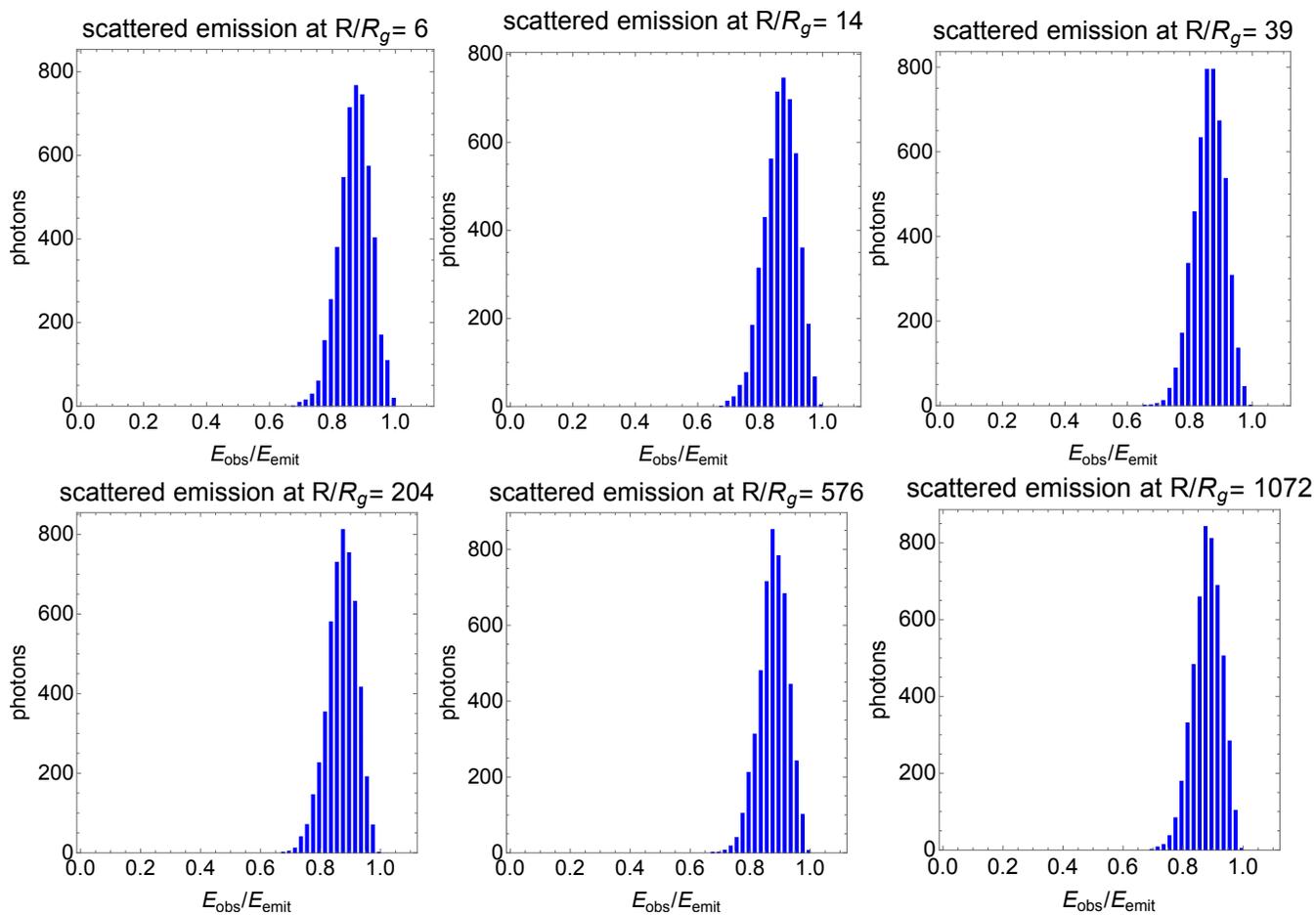

\begin{center}
\includegraphics[trim=0in 0in 0in
0in,keepaspectratio=false,width=2.3in,angle=-0,clip=false]{spec_p1_n1e17_Rg6\_2.pdf}\includegraphics[trim=0in 0in 0in
0in,keepaspectratio=false,width=2.3in,angle=-0,clip=false]{spec_p1_n1e17_Rg14\_2.pdf}\includegraphics[trim=0in 0in 0in
0in,keepaspectratio=false,width=2.3in,angle=-0,clip=false]{spec_p1_n1e17_Rg39\_2.pdf}
\\
\includegraphics[trim=0in 0in 0in
0in,keepaspectratio=false,width=2.3in,angle=-0,clip=false]{spec_p1_n1e17_Rg204\_2.pdf}\includegraphics[trim=0in 0in 0in
0in,keepaspectratio=false,width=2.3in,angle=-0,clip=false]{spec_p1_n1e17_Rg576\_2.pdf}\includegraphics[trim=0in 0in 0in
0in,keepaspectratio=false,width=2.4in,angle=-0,clip=false]{spec_p1_n1e17_Rg1072\_2.pdf}
\end{center}
\caption{MC simulations for scattering of monochromatic emission line spectrum at different emission radius ($R/R_g=6-1072$) through wind for {\tt case 1A} ($p=1$ and $n_{o}=10^{17}$ cm$^{-3}$). {\bf This figure is available as an animation.} }
\label{fig:spec_case1a}
\end{figure}

\section{Results}

\subsection{Scattering of Monochromatic Emission Line}

With the computational scheme  addressed in \S 2, we first investigate a simplistic situation where a packet of photons  with monochromatic energy is emitted from a given radius and scattered by winds of different property. Given the observational implications about wind density and location as discussed in \S 2.1 (see {\bf Fig.~\ref{fig:cont}}), we consider {\it three} regime as listed in {\bf Table~1}; {\tt case 1}: {\it slow} density decay ($p=1$ and $n_o=10^{17}-5 \times 10^{18}$ cm$^{-3}$), {\tt case 2}: {\it moderate} density decay ($p=1.2$ and $n_o=10^{18}-5 \times 10^{19}$ cm$^{-3}$), and {\tt case 3}: {\it fast} density decay ($p=1.7$ and $n_o=10^{19}-5 \times 10^{20}$ cm$^{-3}$), within which different density normalization $n_o$ is examined, respectively; i.e. {\tt case 1A}-{\tt 1C}, {\tt case 2A}-{\tt 2C} and {\tt case 3A}-{\tt 3C}.    

A total number of up to $\sim 10^4$ photons are generally considered in  our multi-dimensional MC simulations for a given radius.  Understanding that a wind of higher optical depth (i.e. higher density) tends to cause more frequent scattering, it would require a much longer computational time for a given number of total photons in a packet. To accommodate such a disparity in computing time, we adopt a different number of photons\footnote[7]{For the purpose of  efficient computations, we have tested and set a minimum number of photons in each case in order to robustly secure a characteristic distribution of photons (i.e. spectra) with little numerical fluctuation. We have thus verified that increasing the size of a photon packet beyond the adopted values here would not change the outcome at a fundamental level.} according to  wind's optical depth in each case as listed in {\bf Table~\ref{tab:tab1}}. As expected, a different choice of the total number of photons in MC calculations lead to a different (absolute) normalization of line profile.  For self-consistency,  we have checked and verified that the present choice of the number of photons is sufficient to correctly characterize a qualitative spectral feature due to scattering process in each case.

When stratified density  of a large-scale wind is slowly falling off with distance for $p=1$ as in {\tt case 1}, the scattering probability with wind optical depth $\tau$ becomes almost constant over every decade in radius, meaning that  the degree of scattering remains roughly the same regardless of from what radius photons are emitted. Therefore, when $n_o$ is relatively low,  many photons manage to travel through a tenuous wind easily, and  emission line profile remains relatively narrow and nearly symmetric  due to a fewer number of scattering events as indeed found in {\bf Figure~\ref{fig:spec_case1a}} for {\tt case 1A} ($p=1$ and $n_o=10^{17}$ cm$^{-3}$). {\bf This figure is available as an animation.} The narrow line shape is thus nearly identical to each other regardless of radius.

\begin{figure}[t]
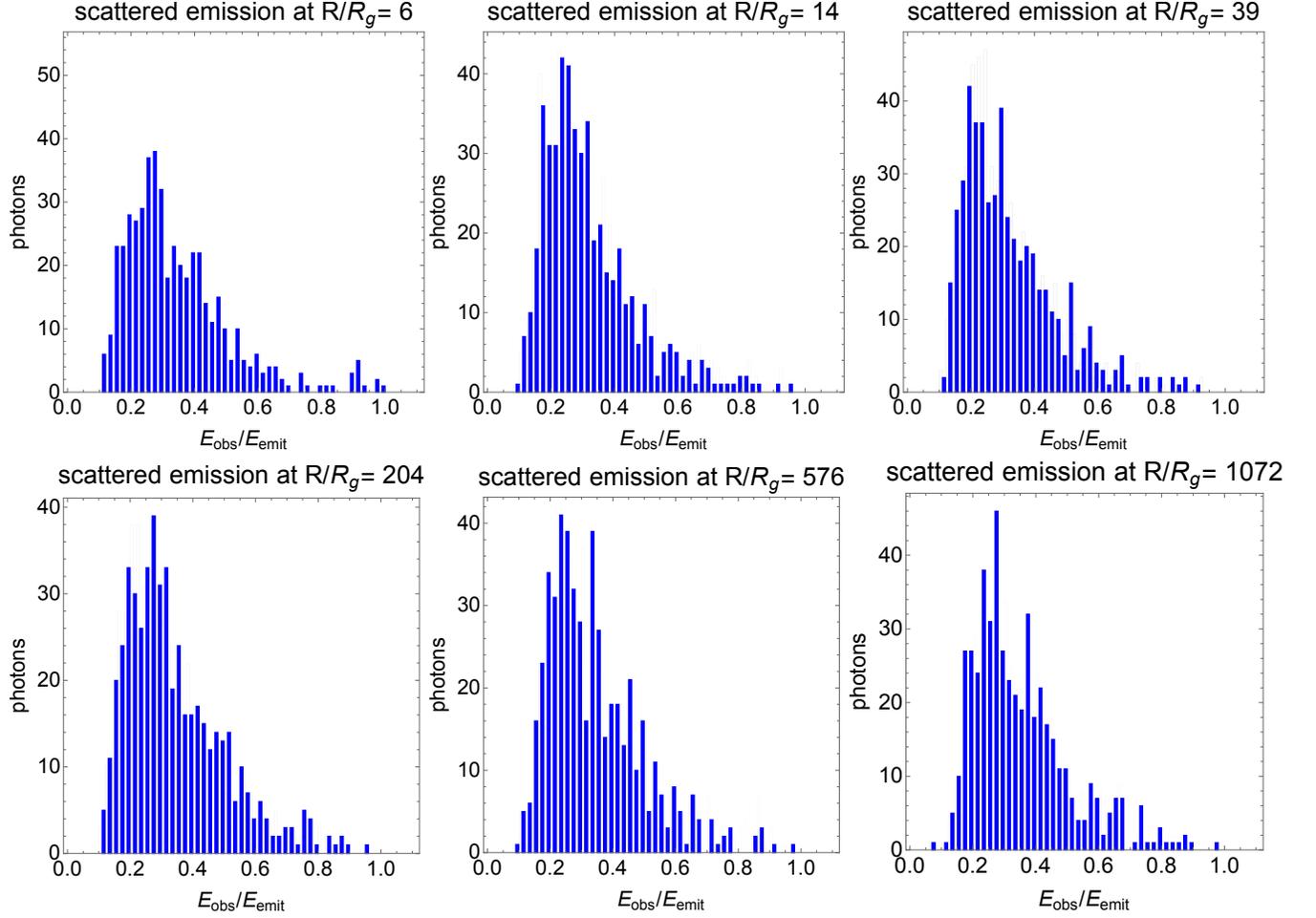

\begin{center}
\includegraphics[trim=0in 0in 0in
0in,keepaspectratio=false,width=2.3in,angle=-0,clip=false]{spec_p1_n5e18_Rg6\_2.pdf}\includegraphics[trim=0in 0in 0in
0in,keepaspectratio=false,width=2.3in,angle=-0,clip=false]{spec_p1_n5e18_Rg14\_2.pdf}\includegraphics[trim=0in 0in 0in
0in,keepaspectratio=false,width=2.3in,angle=-0,clip=false]{spec_p1_n5e18_Rg39\_2.pdf}
\\
\includegraphics[trim=0in 0in 0in
0in,keepaspectratio=false,width=2.3in,angle=-0,clip=false]{spec_p1_n5e18_Rg204\_2.pdf}\includegraphics[trim=0in 0in 0in
0in,keepaspectratio=false,width=2.3in,angle=-0,clip=false]{spec_p1_n5e18_Rg576\_2.pdf}\includegraphics[trim=0in 0in 0in
0in,keepaspectratio=false,width=2.4in,angle=-0,clip=false]{spec_p1_n5e18_Rg1072\_2.pdf}
\end{center}
\caption{Similar to Figure~\ref{fig:spec_case1a} but for {\tt case 1C} ($p=1$ and $n_{o}=5 \times 10^{18}$ cm$^{-3}$). }
\label{fig:spec_case1c}
\end{figure}

On the other hand, when the wind is dense enough with higher $n_o$, almost all the photons emitted from every radius suffer from multiple scatterings before they escape winds. {\bf Figure~\ref{fig:spec_case1c}} demonstrates {\tt case 1C} ($p=1$ and $n_o=5 \times 10^{18}$ cm$^{-3}$).  
Since  photons are overall significantly down-scattered and loose energy, line profile becomes extremely skewed predominantly with lower energy photons as a consequence. 
Note  that we see  more redshifted photons in this case (at lower $E_{\rm obs}/E_{\rm emit}$) independent of radius because the local  optical depth per decade in radius remains the same; i.e. $\Delta \tau \propto \int_{\rm{per~decade}} \tau dr \sim \rm{constant}$. For this reason, the line peak energy appears to be almost unchanged  and the line profile is always asymmetric regardless of the emission radius.    

\begin{figure}[t]
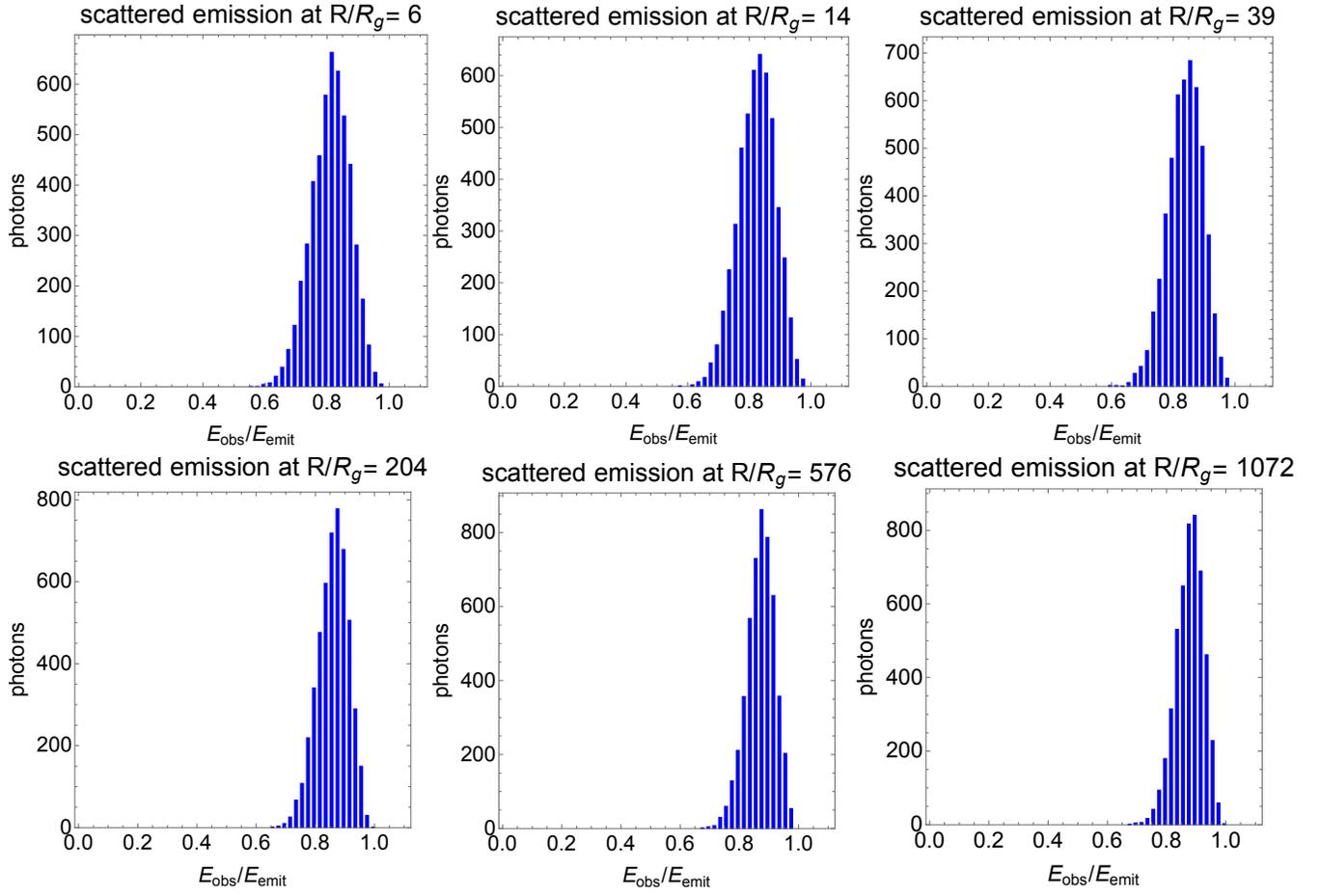

\begin{center}
\includegraphics[trim=0in 0in 0in
0in,keepaspectratio=false,width=2.3in,angle=-0,clip=false]{spec_p12_n1e18_Rg6\_2.pdf}\includegraphics[trim=0in 0in 0in
0in,keepaspectratio=false,width=2.3in,angle=-0,clip=false]{spec_p12_n1e18_Rg14\_2.pdf}\includegraphics[trim=0in 0in 0in
0in,keepaspectratio=false,width=2.3in,angle=-0,clip=false]{spec_p12_n1e18_Rg39\_2.pdf}
\\
\includegraphics[trim=0in 0in 0in
0in,keepaspectratio=false,width=2.3in,angle=-0,clip=false]{spec_p12_n1e18_Rg204\_2.pdf}\includegraphics[trim=0in 0in 0in
0in,keepaspectratio=false,width=2.3in,angle=-0,clip=false]{spec_p12_n1e18_Rg576\_2.pdf}\includegraphics[trim=0in 0in 0in
0in,keepaspectratio=false,width=2.4in,angle=-0,clip=false]{spec_p12_n1e18_Rg1072\_2.pdf}
\end{center}
\caption{Similar to Figure~\ref{fig:spec_case1a} but for {\tt case 2A} ($p=1.2$ and $n_{o}=10^{18}$). }
\label{fig:spec_case2a}
\end{figure}

For winds whose density is more moderately declining with radius as in {\tt case 2}, however, a global effect of scattering is different. To illustrate this point, {\bf Figure~\ref{fig:spec_case2a}} shows how line profile varies with radius for {\tt case 2A} ($p=1.2$ and $n_o=10^{18}$ cm$^{-3}$). They are overall narrow, just like {\tt case 1A} in {\bf Figure~\ref{fig:spec_case1a}},  because the wind is relatively tenuous in density. However, since the wind optical depth is no longer the same per  decade in radius, more photons at larger distances are immune to repeated scattering, allowing  the line width to be distinctively narrower at larger distances (e.g. between $R=6R_g$ and $1072R_g$). This is also clearly understood from the line flux (normalization); i.e. higher flux near $E_{\rm obs}/E_{\rm emit} \sim 1$ for larger radius.

\begin{figure}[t]
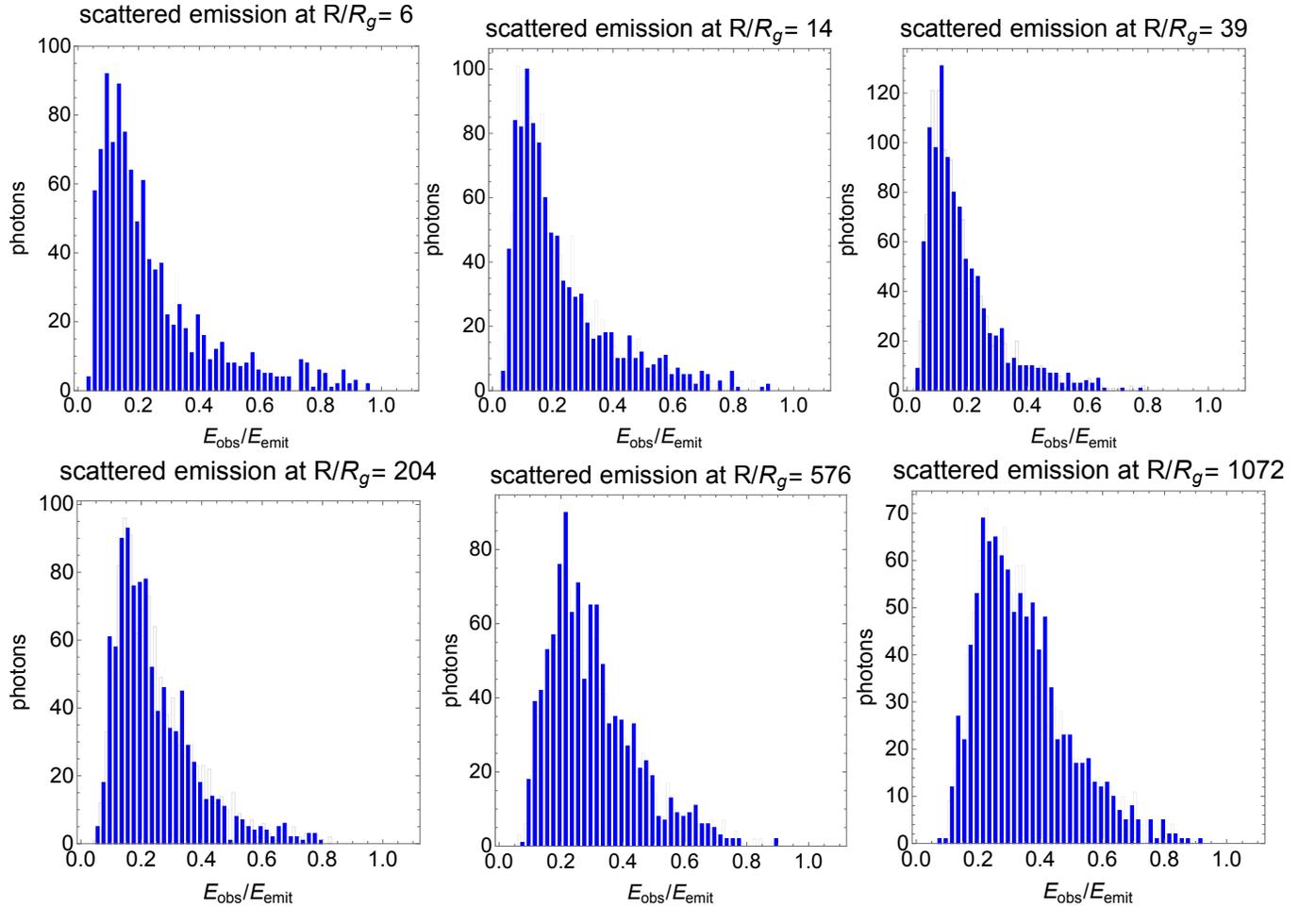

\begin{center}
\includegraphics[trim=0in 0in 0in
0in,keepaspectratio=false,width=2.3in,angle=-0,clip=false]{spec_p12_n5e19_Rg6\_2.pdf}\includegraphics[trim=0in 0in 0in
0in,keepaspectratio=false,width=2.3in,angle=-0,clip=false]{spec_p12_n5e19_Rg14\_2.pdf}\includegraphics[trim=0in 0in 0in
0in,keepaspectratio=false,width=2.3in,angle=-0,clip=false]{spec_p12_n5e19_Rg39\_2.pdf}
\\
\includegraphics[trim=0in 0in 0in
0in,keepaspectratio=false,width=2.4in,angle=-0,clip=false]{spec_p12_n5e19_Rg204\_2.pdf}\includegraphics[trim=0in 0in 0in
0in,keepaspectratio=false,width=2.3in,angle=-0,clip=false]{spec_p12_n5e19_Rg576\_2.pdf}\includegraphics[trim=0in 0in 0in
0in,keepaspectratio=false,width=2.4in,angle=-0,clip=false]{spec_p12_n5e19_Rg1072\_2.pdf}
\end{center}
\caption{Similar to Figure~\ref{fig:spec_case1a}  but for {\tt case 2C} ($p=1.2$ and $n_{o}=5 \times 10^{19}$). {\bf This figure is available as an animation.}  }
\label{fig:spec_case2c}
\end{figure}

This trend is  more pronounced for {\tt case 2C} ($p=1.2$ but $n_o=5 \times 10^{19}$ cm$^{-3}$) in {\bf Figure~\ref{fig:spec_case2c}} where line photons emitted from inner radius are strongly subject to multiple scattering due to higher density, while those emitted from larger radii are not as much. {\bf This figure is available as an animation.} 
The line peak energy clearly shifts towards $E_{\rm obs}/E_{\rm emit} \sim 1$ with increasing radius as a result of less violent scattering process, and the line shape is turned into less asymmetric.

For another comparison, the line profile becomes even more radial-dependent  when the density gradient is much steeper as in {\tt case 3C} ($p=1.7$ and $n_o=5 \times 10^{20}$ cm$^{-3}$) in which more gas is driven as winds from inner region while little material is launched at larger distances. {\bf Figure~\ref{fig:spec_case3c}} demonstrates that the emergent emission in that situation  is very broad and redshifted at smaller radii similar to those in {\bf Figure~\ref{fig:spec_case2c}}. {\bf This figure is available as an animation.}  In contrast, however,  both the line shape and its peak energy are rapidly varying with increasing radius, manifesting a different scattering process across the disk. That is, the line profile becomes almost completely symmetric at $R=1072R_g$ and the peak energy is smoothly shifting towards  $E_{\rm obs}/E_{\rm emit} \sim 1$. In both {\tt cases 2C} and {\tt 3C}, therefore, the qualitative difference in how disk photons are distinctively scattered will come into play for characterizing exactly how a global disk continuum can be influenced as shall be discussed in \S 3.2-\S 3.3 next.   

In {\bf Appendix A}, we show calculated line profiles for the other cases  for comparison.    

\begin{figure}[t]
\begin{center}
\includegraphics[trim=0in 0in 0in
0in,keepaspectratio=false,width=2.3in,angle=-0,clip=false]{spec_p17_n5e20_Rg6\_2.pdf}\includegraphics[trim=0in 0in 0in
0in,keepaspectratio=false,width=2.3in,angle=-0,clip=false]{spec_p17_n5e20_Rg14\_2.pdf}\includegraphics[trim=0in 0in 0in
0in,keepaspectratio=false,width=2.4in,angle=-0,clip=false]{spec_p17_n5e20_Rg39\_2.pdf}
\\
\includegraphics[trim=0in 0in 0in
0in,keepaspectratio=false,width=2.3in,angle=-0,clip=false]{spec_p17_n5e20_Rg204\_2.pdf}\includegraphics[trim=0in 0in 0in
0in,keepaspectratio=false,width=2.3in,angle=-0,clip=false]{spec_p17_n5e20_Rg576\_2.pdf}\includegraphics[trim=0in 0in 0in
0in,keepaspectratio=false,width=2.5in,angle=-0,clip=false]{spec_p17_n5e20_Rg1072\_2.pdf}
\end{center}
\caption{Similar to Figure~\ref{fig:spec_case1a} but for {\tt case 3C} ($p=1.7$ and $n_{o}=5 \times 10^{20}$ cm$^{-3}$). {\bf This figure is available as an animation.}  }
\label{fig:spec_case3c}
\end{figure}

\subsection{Scattering of Local Disk Emission}

\begin{figure}[t]
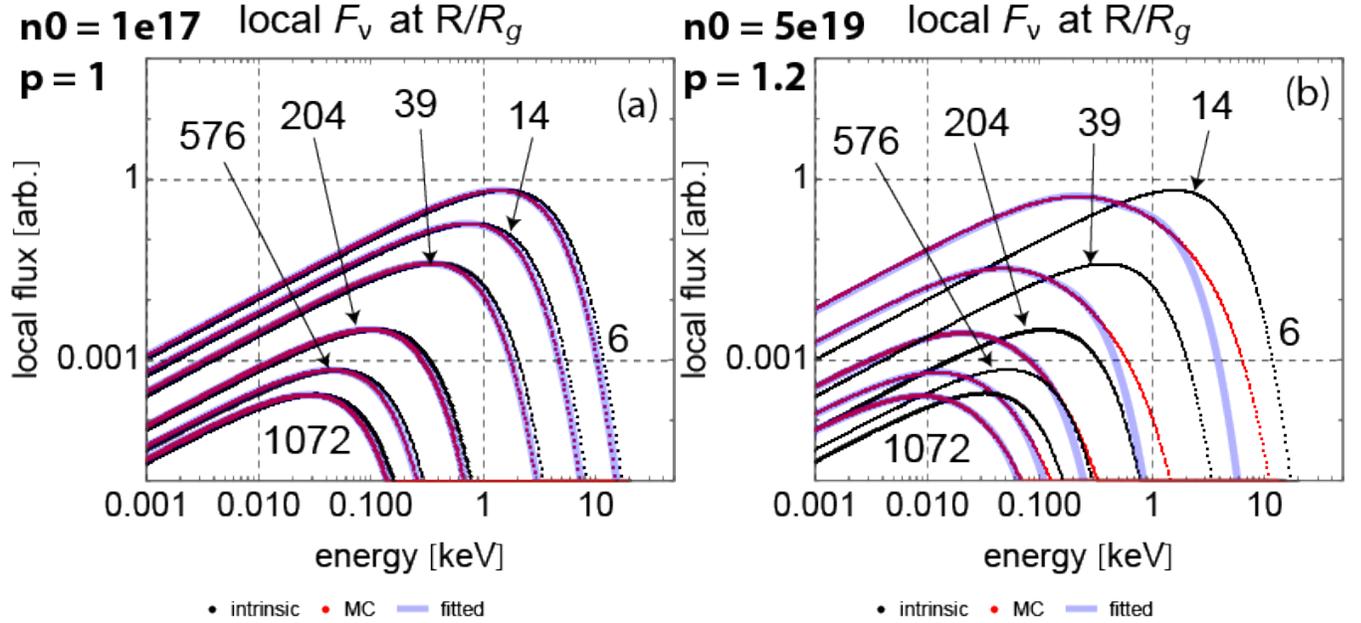

\begin{center}
\includegraphics[trim=0in 0in 0in
0in,keepaspectratio=false,width=3.5in,angle=-0,clip=false]{diskbb_p1_n1e17\_2.png}\includegraphics[trim=0in 0in 0in
0in,keepaspectratio=false,width=3.5in,angle=-0,clip=false]{diskbb_p12_n5e19\_2.png}
%
\end{center}
\caption{Simulated local  thermal spectrum  for various disk radius $R/R_g$ for (a) {\tt case 1A} ($p=1$ and $n_o=10^{17}$ cm$^{-3}$ corresponding to Fig.~\ref{fig:spec_case1a}) and (b) {\tt case 2C} ($p=1.2$ and $n_o=5 \times 10^{19}$ cm$^{-3}$ corresponding to Fig.~\ref{fig:spec_case2c}) showing the intrinsic spectrum in the absence of wind (dark) and Comptonized spectrum in the presence of wind (red). Analytic bestfit with equation~(\ref{eq:bestfit}) is denoted by shaded blue. {\bf These figures are available as animations.}  }
\label{fig:diskbb1}
\end{figure}

With the understanding of how monochromatic photons are scattered in each case, we then further compute a local thermal disk spectrum with winds. In {\bf Figure~\ref{fig:diskbb1}a} we demonstrate how an intrinsic disk continuum (i.e. thermal blackbody; dark) produced locally at different distance will be modified due to scattering (red). Analytic bestfit function expressed in {\bf equation~(\ref{eq:bestfit})}  is shown as well (shaded blue).
In the presence of a  tenuous disk wind as in {\tt case 1A} corresponding to {\bf Figure~\ref{fig:spec_case1a}}, there is only a marginal effect  because the  local radiation is only weakly subject to Compton down-scattering regardless of radius due to low density, and thus the presence of winds is negligible for reshaping the intrinsic spectral shape.  

On the other hand, in higher density regime, local disk emission can be substantially modified by persistent scattering especially for moderately decaying density profile such as {\tt case 2C} corresponding to {\bf Figure~\ref{fig:spec_case2c}}, for example.  It is shown in {\bf Figure~\ref{fig:diskbb1}b} that effective down-scattering is constantly at work over different radius (i.e. low and high in energy) due to a moderately decaying density structure ($p=1.2$), causing a more or less uniform redshift in the scattered spectrum. However, this deviation (i.e. redshift) from the intrinsic spectrum gradually becomes  smaller with increasing radius  as a consequence of  $p=1.2$ profile (e.g. between $R=6R_g$ and $1072R_g$). 
It is seen that the broadband scattered spectrum never converges to the intrinsic one at any radius because of the effective scattering throughout the wind of this type. To further follow up, 
we confirm that the radial  dependence of  the magnitude of down-scattering is clearly evident, for example, for {\tt case 3C} ($p=1.7$ and $n_o=5 \times 10^{20}$ cm$^{-3}$) where little scattering occurs at larger distances due to much lower wind density out there, naturally resulting in a smooth convergence of the low energy tail of the scattered spectrum at large distance (e.g. $R/R_g=1072$ in contrast with smaller radius in {\bf Fig.~\ref{fig:diskbb3}} in Appendix B). See {\bf Appendix B} for the obtained local thermal spectra for the other cases.

\subsection{Scattering of Global MCD Spectra}

\begin{figure}[t]
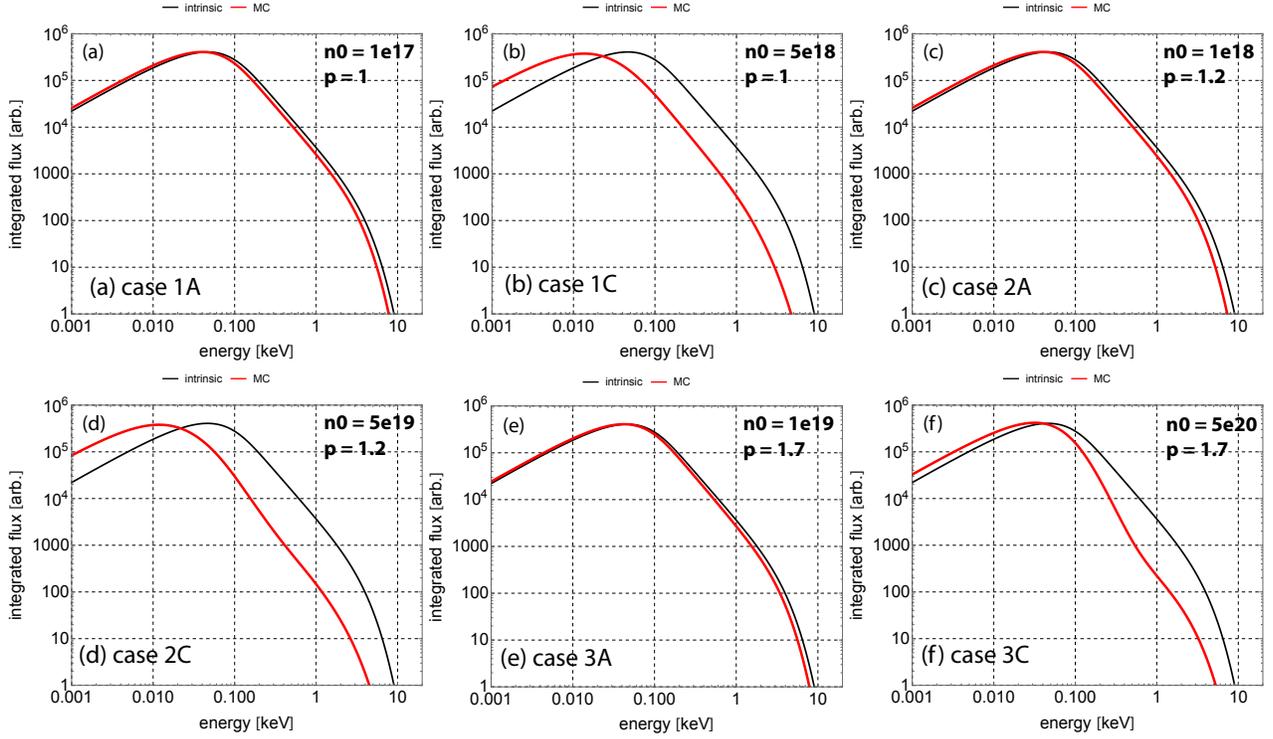

\begin{center}
\includegraphics[trim=0in 0in 0in
0in,keepaspectratio=false,width=2.2in,angle=-0,clip=false]{integrated_diskbb_p1_n1e17\_2.pdf}\includegraphics[trim=0in 0in 0in
0in,keepaspectratio=false,width=2.2in,angle=-0,clip=false]{integrated_diskbb_p1_n5e18\_2.pdf}\includegraphics[trim=0in 0in 0in
0in,keepaspectratio=false,width=2.2in,angle=-0,clip=false]{integrated_diskbb_p12_n1e18\_2.pdf}
\\
\includegraphics[trim=0in 0in 0in
0in,keepaspectratio=false,width=2.2in,angle=-0,clip=false]{integrated_diskbb_p12_n5e19\_2.pdf}\includegraphics[trim=0in 0in 0in
0in,keepaspectratio=false,width=2.2in,angle=-0,clip=false]{integrated_diskbb_p17_n1e19\_2.pdf}\includegraphics[trim=0in 0in 0in
0in,keepaspectratio=false,width=2.2in,angle=-0,clip=false]{integrated_diskbb_p17_n5e20\_2.pdf}
\end{center}
\caption{Simulated MCD spectrum integrated over a large extent of disk surface (i.e. $R/R_g=6-1072$) with winds (red) and without winds (dark) for (a) {\tt case 1A}, (b) {\tt case 1C}, (c) {\tt case 2A}, (d) {\tt case 2C}, (e) {\tt case 3A}, and (f) {\tt case 3C}. }
\label{fig:MCD}
\end{figure}

Given a substantial modification of a locally emitted disk emission in some cases as shown in \S 3.2, we proceed to determine the corresponding MCD spectrum $S_E^{\rm scat}$ in each case. In {\bf Figure~\ref{fig:MCD}} we show a selected gallery of the predicted MCD spectra (red) for (a) {\tt case 1A}, (b) {\tt case 1C}, (c) {\tt case 2A}, (d) {\tt case 2C}, (e) {\tt case 3A} and (f) {\tt case 3C}, in comparison with the intrinsic MCD spectrum (dark). 

We note immediately a possible degeneracy of the scattering effect from different wind profiles; e.g. {\tt case 1A} , {\tt case 2A} and {\tt case 3A}. They are all similar to each other with little deviation from the intrinsic ones because  scattering plays only a partial role in those situations where a large part of the extended wind possesses an insufficient gas density needed to produce a noticeable effect of scattering in the spectrum; i.e. overall low density in {\tt case 1A}, a moderate density decay of initially moderate density in {\tt case 2A} and a rapid declining of initially very high density in {\tt case 3A}, all of which leave little gas over most of the spatial region  to contribute to scattering. Only the innermost wind layer can participate in effective scattering of the inner disk photons of higher energy, which is the reason why only a marginal modification is visible at higher end of the MCD spectrum in these three cases.      

There exists another type of degeneracy; e.g. {\tt case 1C} and {\tt case 2C}. They both exhibit   a semi-uniform deviation from the intrinsic spectrum throughout energy as a result of a persistent scattering over an extended distance almost equally due to less steep density profile of $p=1-1.2$. The whole disk continuum is therefore  redshifted almost systematically throughout energy. Again, this is understood by the fact that in both cases a sufficient amount of wind material is available over a large range of radius, allowing a repeated scattering to constantly occur and continue.  

On the other hand, when the wind of initially higher density  quickly runs out of density with increasing radius as in {\tt case 3C} with $p=1.7$ profile, we see a tangible difference as expected. The innermost disk photons (of higher energy) are substantially scattered by very high wind density, whereas the outer disk photons (of lower energy) can relatively easily  escape the  wind without too much  scattering because of low density out there. The overall MCD spectrum transverse across the entire wind is hence uniquely shaped in a distinct form as seen in which the higher tail (attributed to high energy disk photons from smaller radii) greatly deviates from the intrinsic one, while the lower tail is not so much different from that of the intrinsic one. This type can be classified into another degenerate spectra that is uniquely distinct from the other two types of degenerate spectra as discussed above. To support our view, other MCD spectra are given in {\bf Figure~\ref{fig:MCD2}} in {\bf {Appendix C}} as a supplemental material.

\begin{figure}[t]
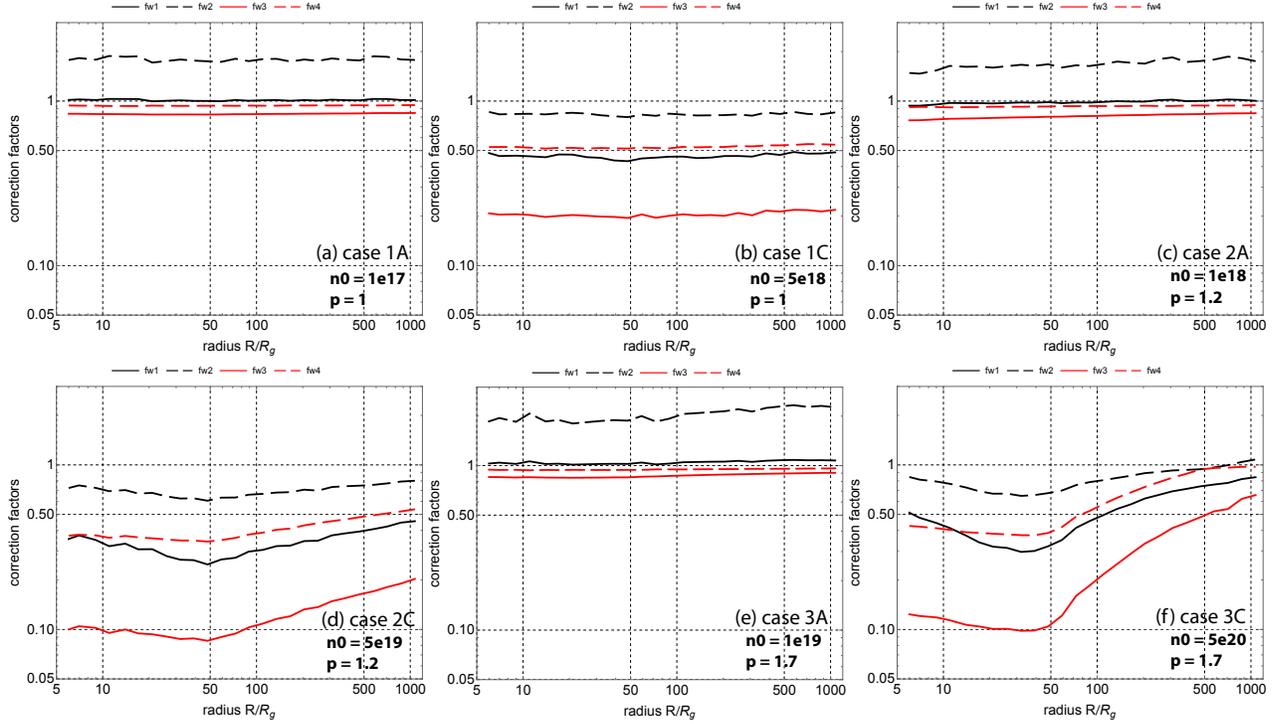

\begin{center}
\includegraphics[trim=0in 0in 0in
0in,keepaspectratio=false,width=2.2in,angle=-0,clip=false]{fw_p1_n1e17\_2.pdf}\includegraphics[trim=0in 0in 0in
0in,keepaspectratio=false,width=2.2in,angle=-0,clip=false]{fw_p1_n5e18\_2.pdf}\includegraphics[trim=0in 0in 0in
0in,keepaspectratio=false,width=2.2in,angle=-0,clip=false]{fw_p12_n1e18\_2.pdf}
\\
\includegraphics[trim=0in 0in 0in
0in,keepaspectratio=false,width=2.2in,angle=-0,clip=false]{fw_p12_n5e19\_2.pdf}\includegraphics[trim=0in 0in 0in
0in,keepaspectratio=false,width=2.2in,angle=-0,clip=false]{fw_p17_n1e19\_2.pdf}\includegraphics[trim=0in 0in 0in
0in,keepaspectratio=false,width=2.2in,angle=-0,clip=false]{fw_p17_n5e20\_2.pdf}
\end{center}
\caption{The bestfit correction factors, $f_{w,1}$ (dark), $f_{w,2}$ (dashed dark), $f_{w,3}$ (solid red), $f_{w,4}$ (dashed red), to approximately characterize the simulated local disk spectra as a function of disk radius ($R/R_g=6-1072$) corresponding to {\bf Figure~\ref{fig:MCD}}.
In the absence of wind, there should be no scattering where $f_{w,i}=1$ $(i=1,2,3,4)$ at all radii. }
\label{fig:fw}
\end{figure}

In order to explicitly quantify the range of deviation from the intrinsic MCD spectrum, we evaluate the evolution  of the bestfit correction factors $f_{w,i}$ ($i=1,2,3,4$) in equation~(\ref{eq:bestfit}) as a function of disk radius $R$ in {\bf Figure~\ref{fig:fw}} corresponding to {\bf Figure~\ref{fig:MCD}}. The other cases are shown in {\bf Figure~\ref{fig:fw2}} in {\bf Appendix C} for completeness. 

Setting aside a small fluctuation due to fitting uncertainties from numerical computation in the bestfit values (where $f_{w,i}=1$ for no scattering), it is clearly found that the range of the factors over radius can be very large when scattering effect varies a lot as a function of radius. Since a different part of the global wind responds differently to incoming disk photons, the obtained spectral deviation is not only wind density dependent (i.e. $n_o$), but also radial dependent as well (i.e. $p$ and $R$). 
Typically, $f_{w,i}$ tends to be smaller at smaller radius reflecting the fact that wind scattering is stronger in consistence with the earlier spectral calculations. In principle,  $f_{w,i}$  converges to unity towards larger radius if scattering effect becomes weaker. In the case of low density regime (e.g. {\tt case 1A}, {\tt case 2A} and {\tt case 3A}), the deviation is very small everywhere over the disk radius as expected.  
Specifically within the current framework, we obtain a vast range of the numerical value for each factor; i.e. $0.3 \lsim f_{w,1} \lsim 1$ (solid dark), $0.7 \lsim f_{w,2} \lsim 1.8$ (dashed dark), $0.1 \lsim f_{w,3} \lsim 0.8$ (solid red), and $0.4 \lsim f_{w,4} \lsim 1$  (dashed red). Due to a highly variable nature of these factors, we find it very hard to unify and uniquely generalize the overall trend across the entire parameter space; i.e. different cases. 
%

\section{Summary \& Discussion}

We have argued in \S 3, by presenting a series of predicted spectra, that powerful accretion disk winds have plausible potential to make a tangible impact on the intrinsic thermal continuum radiation, which is relevant in a specific condition such as disk-dominant high/soft accretion state in BH XRBs. Our  model, while simplistic with only two parameters ($1 \lsim p \lsim 1.7$ and $10^{17} \lsim n_o \rm{[cm^{-3}]} \lsim 10^{20}$), is motivated observationally based on an extensive X-ray observations of blueshifed absorbers detected in BH transients when the disk emission is active as mass-accretion rate is relatively high. We have extended and exploited the existing disk wind model of a stratified structure (in density and velocity) by incorporating multi-dimensional Monte Carlo calculations to address Compton down-scattering of disk photons. Although disk winds in a certain condition tend to extricate the intrinsic MCD spectrum from being changed much, we find that powerful winds in general are capable of substantially modifying the intrinsic disk continuum in a way consistent with the  absorber's density ($n \sim 10^{14-16}$ cm$^{-3}$) and distance ($r/R_g \sim 100-1,000$) range observationally inferred from  state-of-the-art X-ray data of blueshifted absorbers.

Our multi-dimensional MC calculations show that the expected MCD spectrum is phenomenologically (and analytically) better described  by  a superposition of two independent blackbody forms with 4 correction factors. In the parameter space considered in this framework, the range of each factor is respectively found as follows; $0.3 \lsim f_{w,1} \lsim 1.0$, $0.7 \lsim f_{w,2} \lsim 1.8$, $0.1 \lsim f_{w,3} \lsim 0.8$, and $0.4 \lsim f_{w,4} \lsim 1.0$.  Among different scattering results for various wind properties, we have identified 3 distinct types of Comptonized MCD spectra. They can be quantitatively well depicted by a set of average values $(\bar{f}_{w,1}, \bar{f}_{w,2}, \bar{f}_{w,3}, \bar{f}_{w,4})$; i.e. (i) (1.0, 1.8, 0.8, 0.9) for {\it weak/little scattering}, (ii) (0.3, 0.7, 0.1, 0.4) for  {\it semi-uniform scattering} and (iii) (0.5, 0.8, 0.3, 0.6) for {\it central scattering}, as summarized in {\bf Table~\ref{tab:tab2}}. For an illustrative purpose in comparison with the hardening factor $f_c$ by ST95, we also compute the grand average values of $f_{w,1}+f_{w,2}$ and $f_{w,3}+f_{w,4}$, respectively, to get a sense of the overall effect. It is shown that (i) $f_{w,1+2} \simeq  1.4$ and $f_{w,3+4} \simeq 0.9$ for {\it weak/little scattering}, (ii) $f_{w,1+2} \simeq 1.0$ and $f_{w,3+4} \simeq 0.3$ for {\it semi-uniform scattering} and (iii) $f_{w,1+2} \simeq  0.7$ and $f_{w,3+4} \simeq 0.5$ for {\it central scattering}.
Putting this in perspective, the observed disk emission spectrum during high/soft state can thus be extremely down-scattered by factor of up to 70\% relative to the intrinsic one (as in {\tt case 1C}) independent of the hardening effect in the disk atmosphere. 

Our findings should be addressed in tandem with the well-known spectral hardening of thermal continuum due to Compton up-scattering within the well-conceived disk atmosphere in BH XRBs (e.g. ST95; \citealt{Davis05}; \citealt{Davis19}). In the seminal work by ST95, it is shown that the thermal continuum spectrum ought to be modified as a result such that the MCD spectrum coming out of such a region is characterized by a color correction factor of $f_c \sim 1.7-2.0$ for a sensible parameter space of BH XRBs such as LMC~X-3. 
We further feel that a fiducial treatment of spectral hardening with $f_c \sim 1.7-2$ from ST95 may be {\it insufficient} to fully account for the observed disk continuum in the presence of strong wind scattering as demonstrated here.  
It should be reminded again that the introduction of the scattering correction factors $f_{w,i}$ is merely phenomenological to better capture the mathematical form of the scattered MCD spectrum from a mathematical standpoint, and hence there is little physical meaning as opposed to the color correction (hardening) factor of $f_c$ discussed by ST95. Nonetheless, the scattering factors $f_{w,i}$ derived in this work successfully serve as a clear proxy to demonstrate the (extreme) deviation from the standard MCD emission as an inevitable consequence from  wind scattering.

\begin{deluxetable}{cl||l|c|c|c|c||c|c|c} [t]
\tabletypesize{\small} \tablecaption{Average Correction Factors from Wind Scattering in 3 Distinct Regime} \tablewidth{0pt}
\tablehead{Type  & & Type/Example  & $\bar{f}_{w,1}$ & $\bar{f}_{w,2}$ &  $\bar{f}_{w,3}$ &  $\bar{f}_{w,4}$  & $\bar{f}_{w,1+2}$ &  $\bar{f}_{w,3+4}$  & Fig.  } 
\startdata
\hline 
{\bf i} & Weak Scattering & case 1A & 1.0   & 1.8 & 0.8 & 0.9 & 1.4 & 0.9 &  \ref{fig:fw}a   \\
 \hline
{\bf ii} & Semi-Uniform Scattering & case 2C   &  0.3 & 0.7 & 0.1 & 0.4 & 1.0 & 0.3 & \ref{fig:fw}d  \\
  \hline
{\bf iii} & Centeral Scattering & case 3C   & 0.5 & 0.8 & 0.3 & 0.6 & 0.7 & 0.5 & \ref{fig:fw}f  \\
\hline  \label{tab:tab2}
\enddata
\end{deluxetable}

%
In summary, we claim that the presence of  disk winds, especially when the disk emission dominates (as in high/soft state) over nonthermal component, can play a non-negligible role in  reshaping the intrinsic thermal continuum from the disk in an observationally noticeable manner. 
Although our calculations in this paper ignore the effect of the atmosphere, 
it is quite conceivable that disk-driven down-scattering effect (with $f_{w,i}$) can partially (if not substantially) reduce (or counteract)  the effect of atmosphere-induced spectral hardening (for which $f_c \sim 1.7-2$) to the extent that our estimate of the true state of the intrinsic disk radiation might be falsified. Hence, our results necessarily introduce an uncertainty in an accurate estimate of the intrinsic thermal continuum component in real observations, which has been utilized as one of the independent diagnostics for BH spin measurements with the thermal continuum method \citep[e.g.][]{RemillardMcClintock06, Li05,Shafee06,Steiner10, McClintock14} for a population of BH XRBs. Therefore, we caution that the current constraints on BH spin based on the canonical continuum-fitting method may well be ``contaminated" by the effect of strong disk winds. A more rigorous assessment of such a contamination should be performed by global, multi-scale, inflow-outflow simulations.  

We briefly make a note, while beyond the scope of the current work, that the  model considered in this work is essentially mass-invariant; i.e. independent of BH mass. From this perspective, the predicted effect of scattering by disk winds would equally be anticipated from accretion-powered AGNs often exhibiting a broad quasi-thermal feature known as the big blue bump (BBB in optical/UV)  \citep[e.g.][]{Shields78,Malkan83}. In fact, a large portion of Seyfert 1 AGNs and (nearby) quasars have been seen to produce well-known  ionized outflows via UV/X-ray absorption spectroscopy, some of which may well be disk-origin \citep[e.g.][]{CKG03,Yamada24}. 
If AGN disk emission is indeed dominant in the presence of winds as considered in the present work for BH XRBs,  a similar scattering signature can be encoded in the observed BBB, although its statistically significant diagnosis may be challenging due to signal-to-noise ratio and brightness.  

The investigation of Comptonization of radiation around compact objects in fact has a long history in literature. For example, \cite{KHT97} calculated the predicted broadband spectrum of emission from uniform clouds and extended atmospheres with relatively high temperature (e.g. $k_B T_e \sim 50$ keV) and a certain range of optical depth assuming its density profile of $n \propto 1/r$, which somewhat resembles  our models of $p=1$. Their extended cloud spatially spans over $\sim 10^4 R_S$ and it can be identified as disk wind of a stratified geometry within our framework.  While intriguing, their cloud  is much hotter than our photoionized winds (i.e. $k_B T_e \lsim 0.1$ keV), which is the reason why up-scattering is indeed critical in their cloud but negligible in our wind.   

We recognize a few  concerns and caveats in the current model. First, our physical formalism of disk winds  is purely Newtonian \citep[][]{CL94} most notably by  ignoring general relativistic effects in calculating global wind solutions and null ray-tracing with MC approach \citep[e.g.][]{FK07,FK08,Dauser13}. This simplification could mischaracterize the actual physical property of winds (such as density and velocity) at the horizon scale (e.g. tens of \sw radii), but not beyond that region. 

Since our scenario of scattering process is global extending out to $\sim 10^3R_g$ in distance, we are confident that the essential nature of the wind is correctly captured in our calculations. Similarly, null geodesics under strong gravity (including light bending, gravitational redshift, frame-dragging and returning radiation near the BH (e.g. \citealt{FK07,Connors21}, \citealt{Dauser13}, \citealt{Mirzaev24}) should   in principle be considered and treated in a fully relativistic manner. However, from a global perspective of our large-scale wind, the region where these effects practically matter is spatially restricted to a   small central domain near the BH, causing only a marginal influence to the global solution that we obtained in the present framework. 

To quantitatively verify this, we keep track of the fraction  of photons that reach a distant observer (at $R =10^{16}$ cm $\sim 10^9 R_S$ in this case) for {\tt case 1A} ($p=1$ and $n_o=10^{17}$ cm$^{-3}$). It is found that the fraction (makeup) of scattered photons is about 13\% within $R=10R_g$, 40\% within $50R_g$, 56\% within $100R_g$, 87\% within $500R_g$. In other words, about 87\% (=100\%-13\%) of the escaped photons are coming from somewhere beyond  $10R_g$ where general relativistic (GR) effect is insignificant. 
It is evidently true that a ``correct" null trajectory (regardless of scattering), especially in a very small spatial domain near BH,  would be (slightly) different from that shown in Figure~\ref{fig:path} in the presence of GR effect such as light bending. However, what physically matters most in the present work is not necessarily the exact photon trajectories, but scattering process at a global scale (i.e. $R/R_g \lsim 1,000$). 

To follow up more on this point, GR correction to photon energy in \sw geometry is $u_t = \sqrt{1-2/R}$ where $R$ is local radius. Hence, the deviation from a correct observed photon energy will be less than 1\% for $R/R_g  \gsim 100$ while still only 10\% even at $R/R_g=10$. 
It is therefore  reasonable to conceive that the actual scattering distribution (e.g. Figs. 4-8) even with GR effect should not differ too much. 
Furthermore, frame-dragging effect due to BH spin in Kerr geometry would be safely negligible at distance scales that matter for most of the scattering; i.e. well outside the ergosphere, $R/R_g \gg 2$. 

Finally, scattered MCD spectrum is an integral over a larger radial extent of the disk (i.e. $6 \le R/R_g \le 1072$) with factor of $dA = 2 \pi R dR$ that covers a differential surface area. As seen in MCD  spectrum (e.g. Fig.~9), disk emission from smaller radius predominantly contributes to high-energy tail in the total MCD spectrum where GR effect is the strongest but $\int dA$ is the smallest. On the other hand, the bulk of the observed MCD spectrum is attributed to emission from larger distances where GR correction is considered negligible as argued above while $\int dA$ is much larger.  Therefore, we feel it safe to argue that  the innermost disk regions don’t matter to the primary results. Hence, the overall SR/GR correction to the presented MCD spectrum is minimum and qualitatively insignificant to the end results.

We are also aware of a minor caveat concerning how we deal with scattering process. Disk photons locally emitted over different radii (i.e. $R/R_g \simeq 6-1,000$) are all scattered by different part of the extended wind. Our  scheme is focused on simple Compton down-scattering where 
 those photons  at this energy scale (i.e. keV) and wind medium are not  energetic enough to generate additional particles through pair-production or annihilation.
%
Also, our calculations neglect to consider pre-scattered photons at each radius; i.e. we only consider those photons initially emitted from a given radius between $R/R_g=6$ and $1072$. Our  photon packet is hence uniquely initialized per each radius during  MC calculations independent of what's happening at different radii. In other words, no photons initially emitted from a different radius will enter into a different photon packet at a different radius.
This simplicity would pose an issue with accurately handling the secondary photons that come into play for scattering such as, for example, returning radiation of both continuum and scattered photons \citep[e.g.][]{Wilkins20,Connors21,Mirzaev24}, which, however, is beyond the scope of this work. Nonetheless, even without a detailed calculation, one sees that such an additional  process would most likely accelerate more scattering events at smaller radii in favor of further enhancing  down-scattering in the MCD spectrum, but a more quantitative analysis will be necessary for a definitive evaluation. To conclude, we emphasize again that the present work on scattering by winds, which has never been explicitly addressed in literature to date, has a broader  implication that our effort to understand BH XRBs, including the current BH spin measurement with continuum-fitting method, should be carried out with caution.


\begin{acknowledgments}

The initial idea of the present work was partially inspired by a private conversation with Jon Miller and an independent discussion with the participants at ``AGN Winds on the Chesapeake" conference in Easton, MD in 2023. The author acknowledges support from NASA XGS grant (80NSSC23K1021). The author is thankful to an anonymous referee for a number of constructive comments that improved the quality of this manuscript. 

\end{acknowledgments}

\section*{Appendix A: Scattering of Local Emission Line Spectra}

\begin{figure}[t]
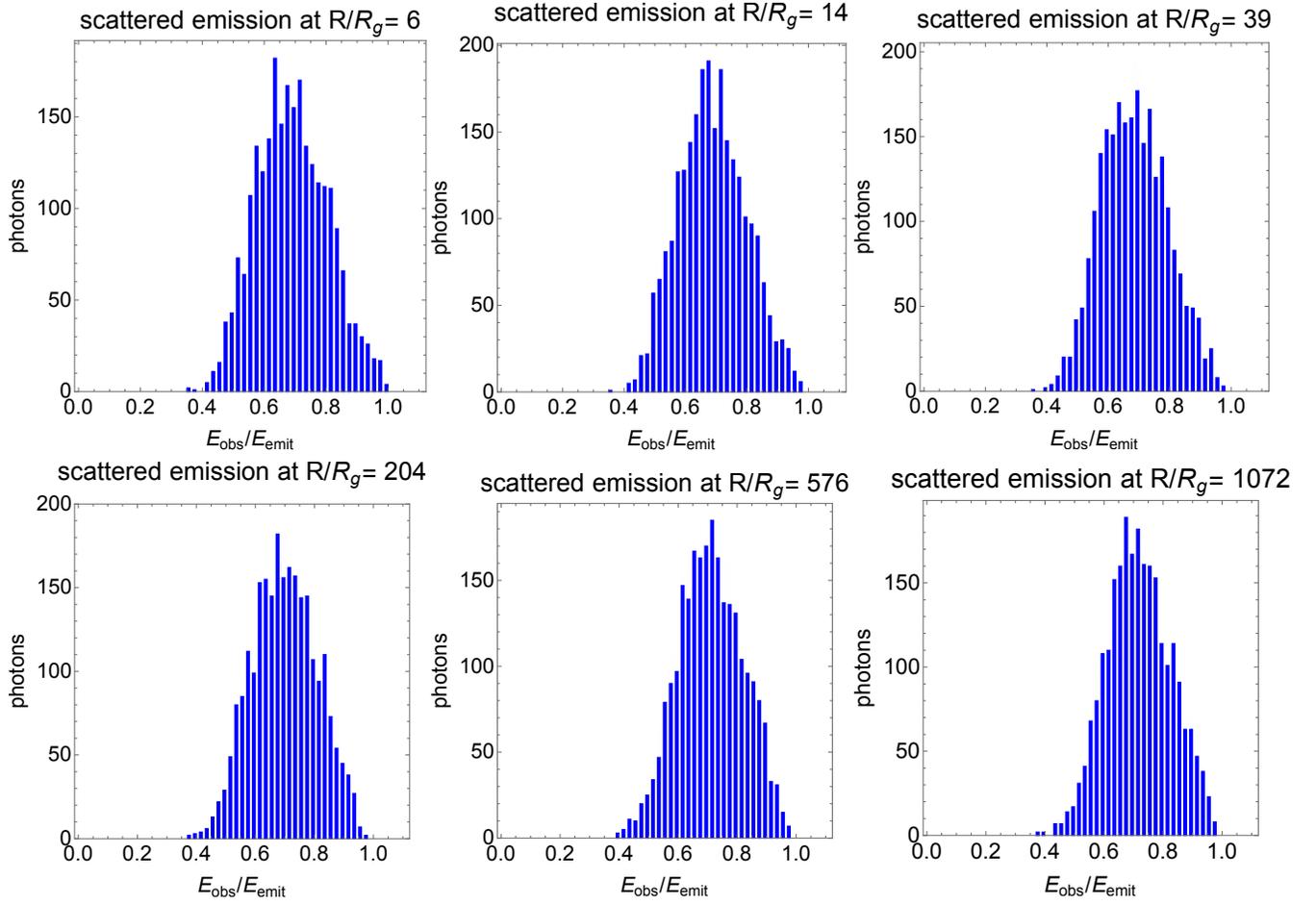

\begin{center}
\includegraphics[trim=0in 0in 0in
0in,keepaspectratio=false,width=2.3in,angle=-0,clip=false]{spec_p1_n1e18_Rg6\_2.pdf}\includegraphics[trim=0in 0in 0in
0in,keepaspectratio=false,width=2.3in,angle=-0,clip=false]{spec_p1_n1e18_Rg14\_2.pdf}\includegraphics[trim=0in 0in 0in
0in,keepaspectratio=false,width=2.3in,angle=-0,clip=false]{spec_p1_n1e18_Rg39\_2.pdf}
\\
\includegraphics[trim=0in 0in 0in
0in,keepaspectratio=false,width=2.3in,angle=-0,clip=false]{spec_p1_n1e18_Rg204\_2.pdf}\includegraphics[trim=0in 0in 0in
0in,keepaspectratio=false,width=2.3in,angle=-0,clip=false]{spec_p1_n1e18_Rg576\_2.pdf}\includegraphics[trim=0in 0in 0in
0in,keepaspectratio=false,width=2.4in,angle=-0,clip=false]{spec_p1_n1e18_Rg1072\_2.pdf}
\end{center}
\caption{Similar to Figure~\ref{fig:spec_case1a} but for {\tt case 1B} ($p=1$ and $n_{o}=10^{18}$ cm$^{-3}$). }
\label{fig:spec_case1b}
\end{figure}

Here, we supplement our discussion on the effect of scattering by disk winds on a monochromatic emission line by providing another representative result for case as shown in Table~\ref{tab:tab1}; i.e. case 1B ({\bf Fig.~\ref{fig:spec_case1b}}), case 2B ({\bf Fig.~\ref{fig:spec_case2b}}), case 3A ({\bf Fig.~\ref{fig:spec_case3a}}), and case 3B ({\bf Fig.~\ref{fig:spec_case3b}}).

\begin{figure}[t]
\begin{center}
\includegraphics[trim=0in 0in 0in
0in,keepaspectratio=false,width=2.3in,angle=-0,clip=false]{spec_p12_n1e19_Rg6\_2.pdf}\includegraphics[trim=0in 0in 0in
0in,keepaspectratio=false,width=2.3in,angle=-0,clip=false]{spec_p12_n1e19_Rg14\_2.pdf}\includegraphics[trim=0in 0in 0in
0in,keepaspectratio=false,width=2.3in,angle=-0,clip=false]{spec_p12_n1e19_Rg39\_2.pdf}
\\
\includegraphics[trim=0in 0in 0in
0in,keepaspectratio=false,width=2.3in,angle=-0,clip=false]{spec_p12_n1e19_Rg204\_2.pdf}\includegraphics[trim=0in 0in 0in
0in,keepaspectratio=false,width=2.3in,angle=-0,clip=false]{spec_p12_n1e19_Rg576\_2.pdf}\includegraphics[trim=0in 0in 0in
0in,keepaspectratio=false,width=2.4in,angle=-0,clip=false]{spec_p12_n1e19_Rg1072\_2.pdf}
\end{center}
\caption{Similar to Figure~\ref{fig:spec_case1a} but for {\tt case 2B} ($p=1.2$ and $n_{o}=10^{19}$ cm$^{-3}$). }
\label{fig:spec_case2b}
\end{figure}

\begin{figure}[t]
\begin{center}
\includegraphics[trim=0in 0in 0in
0in,keepaspectratio=false,width=2.3in,angle=-0,clip=false]{spec_p17_n1e19_Rg6\_2.pdf}\includegraphics[trim=0in 0in 0in
0in,keepaspectratio=false,width=2.3in,angle=-0,clip=false]{spec_p17_n1e19_Rg14\_2.pdf}\includegraphics[trim=0in 0in 0in
0in,keepaspectratio=false,width=2.3in,angle=-0,clip=false]{spec_p17_n1e19_Rg39\_2.pdf}
\\
\includegraphics[trim=0in 0in 0in
0in,keepaspectratio=false,width=2.3in,angle=-0,clip=false]{spec_p17_n1e19_Rg204\_2.pdf}\includegraphics[trim=0in 0in 0in
0in,keepaspectratio=false,width=2.3in,angle=-0,clip=false]{spec_p17_n1e19_Rg576\_2.pdf}\includegraphics[trim=0in 0in 0in
0in,keepaspectratio=false,width=2.4in,angle=-0,clip=false]{spec_p17_n1e19_Rg1072\_2.pdf}
\end{center}
\caption{Similar to Figure~\ref{fig:spec_case1a} but for {\tt case 3A} ($p=1.7$ and $n_{o}=10^{19}$ cm$^{-3}$). }
\label{fig:spec_case3a}
\end{figure}

\begin{figure}[t]
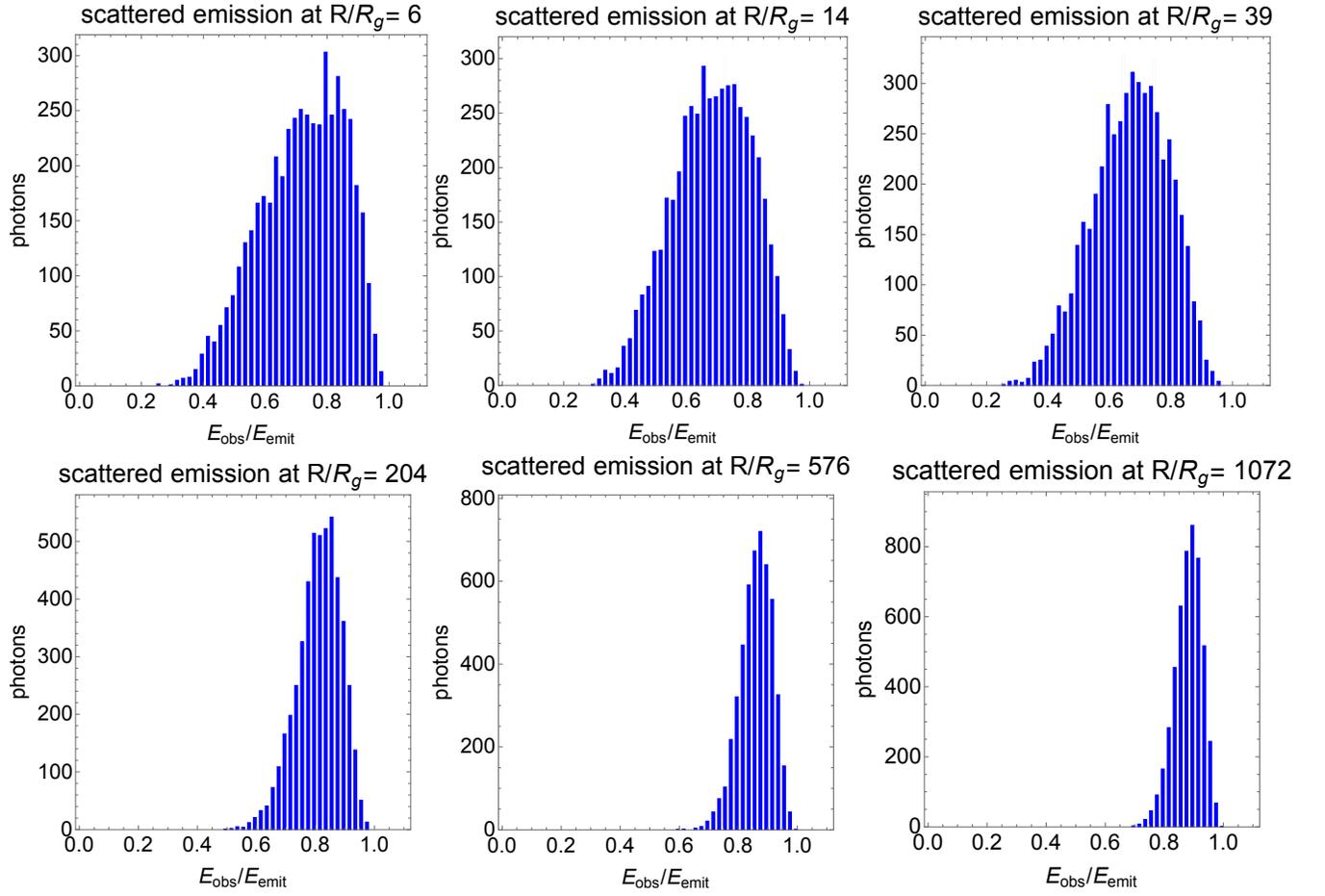

\begin{center}
\includegraphics[trim=0in 0in 0in
0in,keepaspectratio=false,width=2.3in,angle=-0,clip=false]{spec_p17_n1e20_Rg6\_2.pdf}\includegraphics[trim=0in 0in 0in
0in,keepaspectratio=false,width=2.3in,angle=-0,clip=false]{spec_p17_n1e20_Rg14\_2.pdf}\includegraphics[trim=0in 0in 0in
0in,keepaspectratio=false,width=2.3in,angle=-0,clip=false]{spec_p17_n1e20_Rg39\_2.pdf}
\\
\includegraphics[trim=0in 0in 0in
0in,keepaspectratio=false,width=2.3in,angle=-0,clip=false]{spec_p17_n1e20_Rg204\_2.pdf}\includegraphics[trim=0in 0in 0in
0in,keepaspectratio=false,width=2.3in,angle=-0,clip=false]{spec_p17_n1e20_Rg576\_2.pdf}\includegraphics[trim=0in 0in 0in
0in,keepaspectratio=false,width=2.4in,angle=-0,clip=false]{spec_p17_n1e20_Rg1072\_2.pdf}
\end{center}
\caption{Similar to Figure~\ref{fig:spec_case1a} but for {\tt case 3B} ($p=1.7$ and $n_{o}=10^{20}$ cm$^{-3}$). }
\label{fig:spec_case3b}
\end{figure}

\section*{Appendix B: Scattering of Local Disk Emission Spectra }

\begin{figure}[t]
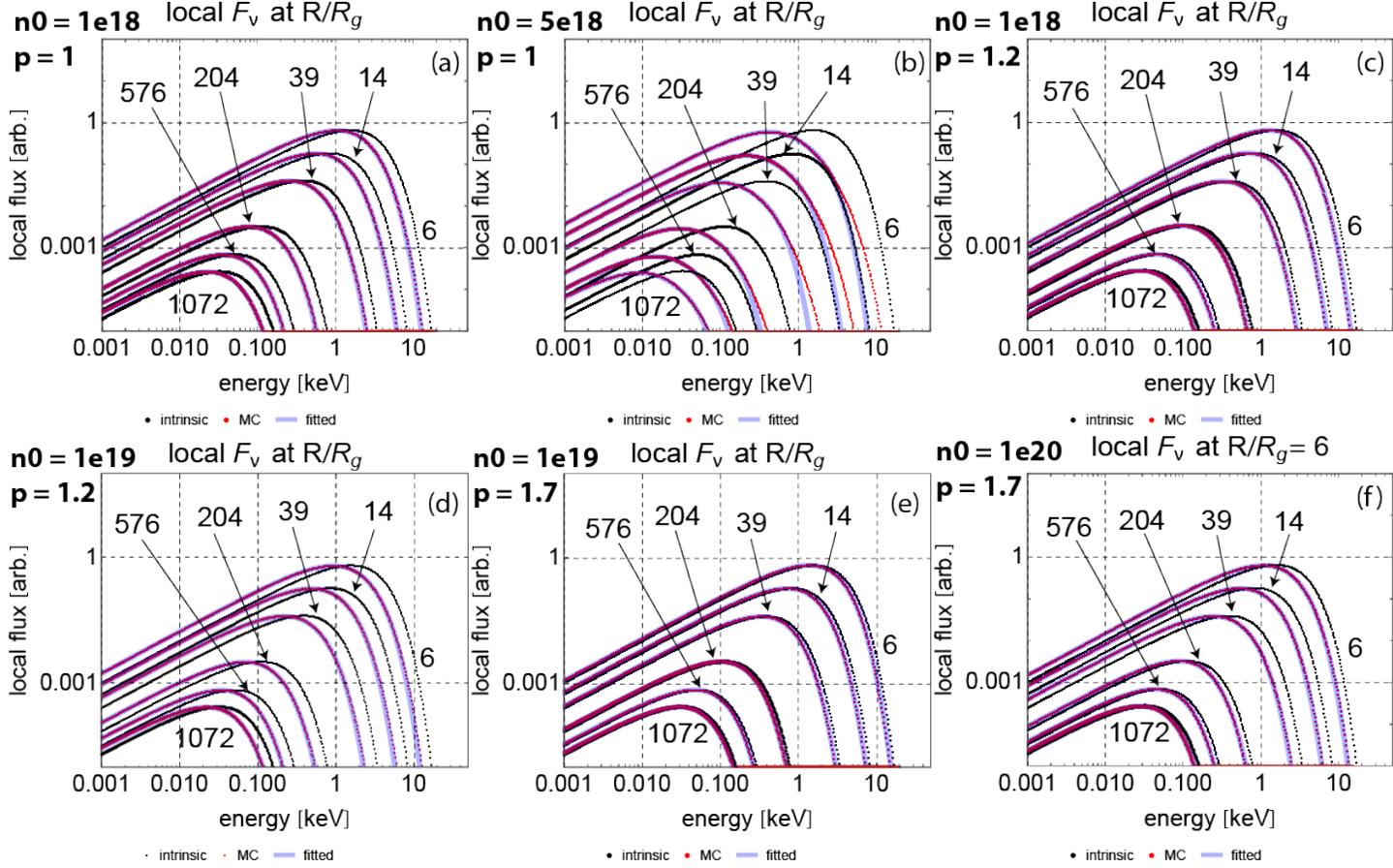

\begin{center}
\includegraphics[trim=0in 0in 0in
0in,keepaspectratio=false,width=2.5in,angle=-0,clip=false]{diskbb_p1_n1e18\_2.png}\includegraphics[trim=0in 0in 0in
0in,keepaspectratio=false,width=2.5in,angle=-0,clip=false]{diskbb_p1_n5e18\_2.png}\includegraphics[trim=0in 0in 0in
0in,keepaspectratio=false,width=2.5in,angle=-0,clip=false]{diskbb_p12_n1e18\_2.png}
\\
\includegraphics[trim=0in 0in 0in
0in,keepaspectratio=false,width=2.5in,angle=-0,clip=false]{diskbb_p12_n1e19\_2.png}\includegraphics[trim=0in 0in 0in
0in,keepaspectratio=false,width=2.5in,angle=-0,clip=false]{diskbb_p17_n1e19\_2.png}\includegraphics[trim=0in 0in 0in
0in,keepaspectratio=false,width=2.5in,angle=-0,clip=false]{diskbb_p17_n1e20\_2.png}
\end{center}
\caption{Similar to Figure~\ref{fig:diskbb1} but for (a) {\tt case 1B}, (b) {\tt case 1C}, (c) {\tt case 2A}, (d) {\tt case 2B}, (e) {\tt case 3A} and (f) {\tt case 3B}. }
\label{fig:diskbb2}
\end{figure}

\begin{figure}[t]
\begin{center}
\includegraphics[trim=0in 0in 0in
0in,keepaspectratio=false,width=2.5in,angle=-0,clip=false]{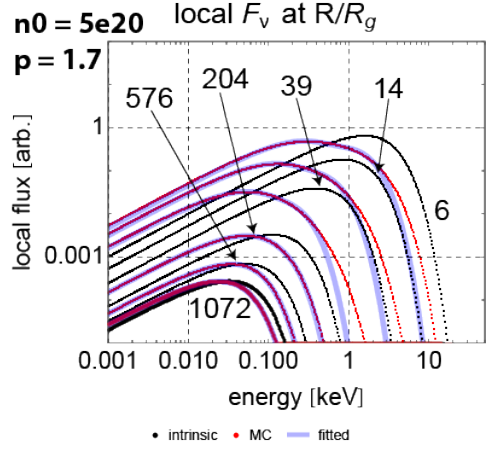}
\end{center}
\caption{Similar to Figure~\ref{fig:diskbb1} but for {\tt case 3C}. {\bf This figure is available as an animation.} }
\label{fig:diskbb3}
\end{figure}

Here, we supplement our discussion on the effect of scattering by disk winds on the local disk emission by providing another representative result for case as shown in Table~\ref{tab:tab1}; i.e. {\tt case 1B}, {\tt case 1C}, {\tt case 2A}, {\tt case 2B}, {\tt case 3A}, and {\tt case 3B}. See {\bf Figures~\ref{fig:diskbb2}-\ref{fig:diskbb3}}. {\bf Figure~\ref{fig:diskbb3} is available as an animation.} 

\section*{Appendix C: Scattering of Global MCD Spectra }

\begin{figure}[t]
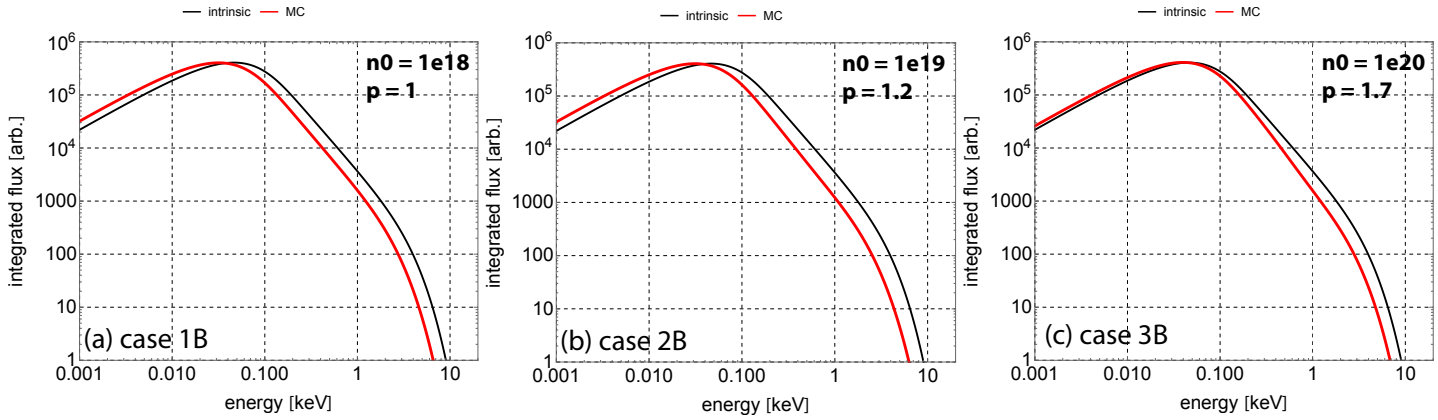

\begin{center}
\includegraphics[trim=0in 0in 0in
0in,keepaspectratio=false,width=2.5in,angle=-0,clip=false]{integrated_diskbb_p1_n1e18\_2.pdf}\includegraphics[trim=0in 0in 0in
0in,keepaspectratio=false,width=2.5in,angle=-0,clip=false]{integrated_diskbb_p12_n1e19\_2.pdf}\includegraphics[trim=0in 0in 0in
0in,keepaspectratio=false,width=2.5in,angle=-0,clip=false]{integrated_diskbb_p17_n1e20\_2.pdf}
\end{center}
\caption{Similar to Figure~\ref{fig:MCD} but for (a) {\tt case 1B}, (b) {\tt case 2B} and (c) {\tt case 3B}. }
\label{fig:MCD2}
\end{figure}

\begin{figure}[t]
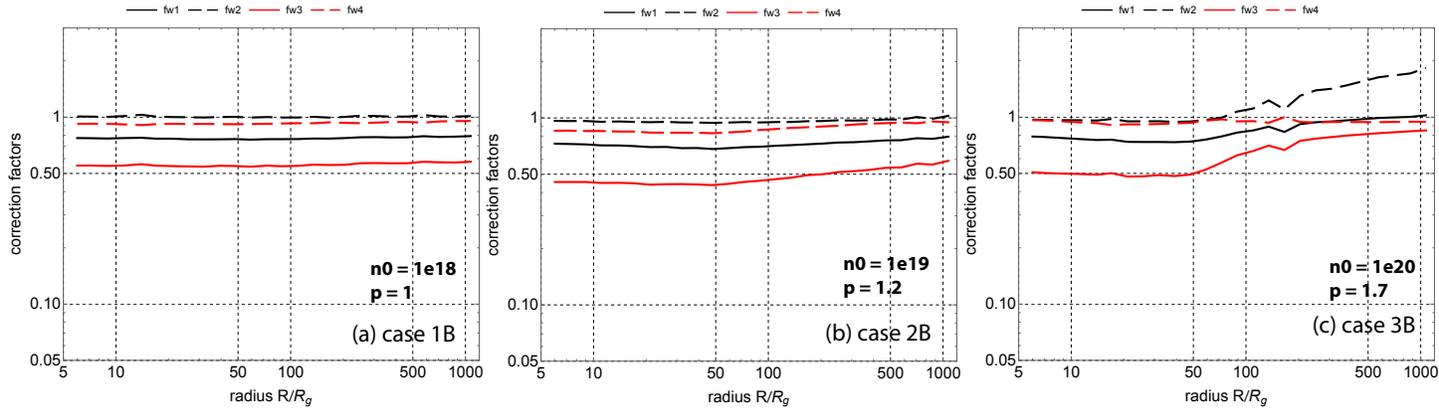

\begin{center}
\includegraphics[trim=0in 0in 0in
0in,keepaspectratio=false,width=2.5in,angle=-0,clip=false]{fw_p1_n1e18\_2.pdf}\includegraphics[trim=0in 0in 0in
0in,keepaspectratio=false,width=2.5in,angle=-0,clip=false]{fw_p12_n1e19\_2.pdf}\includegraphics[trim=0in 0in 0in
0in,keepaspectratio=false,width=2.5in,angle=-0,clip=false]{fw_p17_n1e20\_2.pdf}
\end{center}
\caption{Similar to Figure~\ref{fig:fw} but for (a) case 1B, (b) case 2B and (c) case 3B, respectively  corresponding to Figure~\ref{fig:MCD2}. }
\label{fig:fw2}
\end{figure}

Here, we supplement our discussion on the effect of scattering by disk winds on the global MCD emission  by providing another representative result for case as shown in Table~\ref{tab:tab1}; i.e. case 1B, case 2B, and case 3B ({\bf Fig.~\ref{fig:MCD2}}). The corresponding correction factors, $f_{w,i}$, are also shown ({\bf Fig.~\ref{fig:fw2}}).  

\end{document}